\documentclass[11pt,a4paper]{article}

\usepackage[T1]{fontenc}
\usepackage[utf8]{inputenc}
\usepackage{amsmath,amssymb,amsfonts}   % load math before the font package
\usepackage{mathptmx}                    % Times text + matching math
\usepackage[scaled=0.92]{helvet}         % Helvetica for sans (headings/figures)
\usepackage{microtype}                   % subtle spacing improvements

\usepackage{graphicx}
\usepackage{array}
\usepackage{booktabs}
\usepackage{xcolor}
\usepackage{textcomp}
\usepackage[title]{appendix}

\usepackage{geometry}
\usepackage{titlesec}
\titleformat{\section}{\normalfont\large\bfseries}{\thesection}{0.6em}{}
\titleformat{\subsection}{\normalfont\normalsize\bfseries}{\thesubsection}{0.6em}{}
\titleformat{\subsubsection}[runin]{\normalfont\bfseries}{}{0pt}{}
\titlespacing*{\subsubsection}{0pt}{0.7\baselineskip}{0.5em}
\usepackage[authoryear,round]{natbib}
\usepackage[colorlinks=true,
            linkcolor={blue!75!black},
            citecolor={blue!75!black},
            urlcolor={teal!70!black},
            breaklinks=true]{hyperref}

\newcommand{\ledm}{\mathrm{LE\text{-}DM}}

\begin{document}

% ----------------------------------------------------------------
%  Title block
% ----------------------------------------------------------------
\begin{center}
  {\LARGE\bfseries
   Dynamic masking for boundary-aware velocity reconstruction\\[2pt]
   in volumetric particle tracking with moving solids\par}

  \vspace{1.1em}

  {\large
   Jibu Tom Jose$^{1}$,\;
   Arieh Jacobson$^{1}$,\;
   Dhanush Vittal Shenoy$^{1}$,\;
   Steven H. Frankel$^{1}$,\;
   Omri Ram$^{1,\ast}$\par}

  \vspace{0.7em}

  {\small
   $^{1}$Faculty of Mechanical Engineering,
   Technion -- Israel Institute of Technology,
   Haifa 3200003, Israel\par}

  \vspace{0.25em}

  {\small
   $^{\ast}$Corresponding author:
   \href{mailto:omri.ram@technion.ac.il}{omri.ram@technion.ac.il}\par}

  \vspace{0.5em}

  {\small \today\par}
\end{center}

\vspace{0.6em}

% ----------------------------------------------------------------
%  Abstract and keywords
% ----------------------------------------------------------------
\begin{center}
\begin{minipage}{0.90\textwidth}
{\small
\noindent\textbf{Abstract}\par
\vspace{2pt}
\noindent
Volumetric particle tracking velocimetry (PTV) produces scattered Lagrangian tracks that must be reconstructed on an Eulerian grid before velocity gradients, pressure, or hydrodynamic loads can be evaluated. This step is usually performed on a domain treated as entirely fluid. When a solid body lies within the measurement volume, its surface kinematics are not imposed and the reconstruction is weakest in the steep-gradient region next to the body. We introduce LE-DM (Lagrangian-to-Eulerian reconstruction with Dynamic Masking), a constrained reconstruction framework for moving solid boundaries. A time-dependent signed-distance function classifies grid nodes as open fluid, boundary shell, or solid interior. The particle data, incompressibility constraint, prescribed surface velocity, and regularization terms are then assembled on the masked domain within a single solve. The method requires only a signed-distance field and a surface velocity, allowing stationary walls, translating, rotating, multiple, and deforming bodies to be represented in the same formulation. LE-DM is assessed using an analytical oscillating sphere, synthetic tracks from a CFD rising-sphere simulation, and a refractive-index-matched tomographic-PTV experiment on a freely rising sphere. The surface kinematics are enforced to solver tolerance, while the bulk reconstruction remains unchanged where no body is present. In the analytical case, the first-cell error is reduced from 14\% to 3\% of the body speed. In the experiment, LE-DM recovers the independently measured surface velocity, whereas an all-fluid reconstruction does not. The result is a divergence-free, boundary-consistent velocity field for pressure and force estimation.
\par\vspace{6pt}
}
\end{minipage}
\end{center}

\vspace{1.0em}

% ─────────────────────────────────────────────────────────
\section{Introduction}
\label{sec:introduction}
% ─────────────────────────────────────────────────────────

Time-resolved volumetric particle tracking velocimetry (PTV) is now widely used for measuring three-dimensional unsteady flows. Its development has been enabled by several complementary advances, including tomographic imaging for volumetric particle reconstruction \citep{elsinga2006tomographic}, volume self-calibration for improved multi-camera registration \citep{wieneke2008volume}, iterative particle reconstruction at high seeding densities \citep{wieneke2013iterative}, and Shake-the-Box (STB) for robust Lagrangian particle tracking \citep{schanz2016shake}. Together, these methods allow large numbers of particle trajectories to be recovered and provide direct access to particle position, velocity, and material acceleration in a Lagrangian frame \citep{jahn2021advanced,schroder2023lpt}. The particle tracks themselves, however, are usually an intermediate measurement rather than the final quantity of interest. Physical interpretation often relies on derived Eulerian quantities, including velocity gradients, vorticity, strain rate, pressure, wall loading, and hydrodynamic forces, which must be recovered from the scattered Lagrangian data. The accuracy of this reconstruction propagates through the rest of the measurement chain. Velocity errors are amplified by the spatial differentiation used to obtain gradients and vorticity, and they enter the pressure field through the integration of the material acceleration, so inaccuracies in the near-wall velocity translate directly into errors in the recovered pressure and the resulting force estimates \citep{vanGent2017comparative, sciacchitano2025accuracy}.

Several approaches have been developed for reconstructing Eulerian fields from scattered particle tracks, with the main distinction lying in how sparse and noisy Lagrangian data are regularized. Direct binning and local interpolation provide simple and computationally efficient estimates, but they fit the data locally without enforcing the governing equations. Their accuracy therefore deteriorates when the particle field is sparse and the local track support is insufficient to constrain the velocity field \citep{agarwal2021reconstructing}. Physics-constrained approaches address this limitation by incorporating constraints such as incompressibility, vorticity transport, or smoothness into the reconstruction. Examples include vortex-in-cell and its later developments, VIC+ and VIC\# \citep{schneiders2016dense,scarano2022dense,jeon2022fine}; the B-spline FlowFit  \citep{gesemann2015particle, vanGent2017comparative, godbersen2024flowfit3}; constrained cost minimization (CCM) \citep{agarwal2021reconstructing}; singular-value-decomposition-based interpolation \citep{sheng2008using}; meshless radial-basis-function regression \citep{li2024lagrangian}; and more recent physics-informed or neural data-assimilation methods for sparse tracks \citep{clarkDiLeoni2023reconstructing,zhou2024benchmarking}. Although these methods differ substantially in their numerical formulation, they share a common assumption. The reconstruction domain, including any solid boundary within it, is fixed and specified in advance, so the same fluid-solid partition is used throughout the reconstruction.

This requirement becomes restrictive when a solid body lies within the measurement volume. Such cases arise in flows around fixed obstacles, wall-bounded configurations, sedimenting or rising particles, droplets and bubbles, biological locomotion, and fluid-structure interaction. The body excludes part of the Cartesian grid, the surrounding fluid must satisfy a kinematic condition at its surface, and the near-wall region is often the worst sampled part of the volume because optical occlusion, reflections, particle drop-out, and uncertainty in the detected body position degrade the local track quality. For moving bodies, this difficulty is compounded by the fact that the fluid domain changes in time, as a grid node occupied by the solid at one instant may be part of the open fluid at the next.

Applying an all-fluid reconstruction to such data introduces errors precisely where the reconstructed field is most important. First, incompressibility conditions may be enforced at grid nodes located inside the solid, producing non-physical gradients in a region where no fluid is present. Second, the no-slip or rigid-body condition at the surface is absent, so the reconstructed velocity field may penetrate the body and fail to match the known surface motion. These errors are concentrated near the boundary, where velocity gradients are largest, and can propagate directly into the pressure field, since pressure recovery from PTV is commonly based on the integration of material acceleration \citep{liu2006instantaneous,neeteson2015pressure,vanGent2017comparative,wang2019gpu}. Recent benchmarks of dense particle-track assimilation show the same sensitivity: pressure recovery is generally more reliable in the bulk, whereas surface-pressure estimation near rigid or deformable boundaries remains substantially more challenging \citep{sciacchitano2025accuracy}. A more consistent approach is to include the boundary within the velocity reconstruction itself, rather than treating it only as a visual mask or as a post-processing correction.

Previous studies have addressed solid boundaries at two stages of the measurement chain. At the image-to-track stage, visual-hull and silhouette masks exclude the solid region from the reconstructed intensity volume and reduce ghost particles caused by optical occlusion \citep{adhikari2012visual}. Object-aware Lagrangian particle tracking (OA-LPT) extends this approach by representing a known object through per-camera depth maps within the iterative particle reconstruction and self-calibration steps of STB. This improves particle recovery when the line of sight is partly blocked \citep{wieneke2024lagrangian}. These methods improve the particle tracks, but they do not modify the subsequent Eulerian reconstruction. Solid boundaries have also been incorporated directly at the reconstruction stage. Constrained meshless methods can impose no slip on a fixed boundary as a hard constraint in a least-squares regression, as demonstrated for a fixed wall and a stationary Stokes sphere \citep{sperotto2022meshless,sperotto2024spicy}. In vorticity-based assimilation, \citet{cakir2022dense} introduced boundary handling into VIC+ using two fluid-structure-interaction formulations. The first conforms a deforming body-fitted grid to the moving surface, and the second retains a uniform grid while imposing the surface condition through surface source and doublet (panel-method) singularities on an immersed boundary. These formulations were demonstrated for a fixed periodic-hill geometry and an in-place deforming membrane. 

What these reconstruction-stage methods share is that the solid boundary either occupies a fixed region of space or deforms about a prescribed location, as in a stationary wall, a fixed obstacle, or an in-place flexible surface. The fluid-solid partition is either static or confined to a local region around a fixed mean position, allowing the boundary treatment to be built around a prescribed configuration. A body translating through the measurement volume presents a different problem. The occupied region moves across the reconstruction grid, so fluid and solid nodes must be reclassified in time, and a node that is part of the bulk fluid at one instant may have been inside the body shortly before. Recent comparative assessments of dense assimilation algorithms reflect the former scope, considering fluid-structure cases in which a wall or flexible panel oscillates at a fixed location \citep{sciacchitano2025accuracy}. In addition, the boundary is usually introduced through method-specific machinery, such as a body-fitted mesh regenerated during the solution procedure or a singularity formulation tied to a vorticity-transport solver. The methods of \citet{cakir2022dense} address deforming boundaries, but in both cases the surface remains at a fixed mean location. ALE-VIC+ moves a body-fitted mesh with the surface rather than retaining a fixed grid, and the immersed variant is demonstrated only for in-place deformation. Neither reclassifies fixed-grid nodes as a body sweeps through the volume, and neither corrects a temporal prior at newly exposed fluid nodes. A general treatment of a body translating through the measurement volume, with fluid and solid nodes reclassified as the body moves, within a single grid-based reconstruction has not yet been established.

This study introduces LE-DM, a Lagrangian-to-Eulerian reconstruction with Dynamic Masking, designed for flows with moving boundaries. The method builds on the constrained cost-minimization framework of \citet{agarwal2021reconstructing}, retaining the enforcement of physical constraints as hard conditions within a single solve, while extending the formulation to domains in which the fluid and solid regions evolve in time. At each time step, a signed-distance function classifies every grid node as open fluid, boundary shell, or solid interior. Incompressibility is enforced only in the fluid. Nodes in the boundary shell and solid interior are constrained to the prescribed boundary velocity, and one-sided divergence stencils prevent the incompressibility constraint from crossing the surface. Three additional terms account for body motion and the use of experimental measurements. The data-fidelity term is damped near the wall according to the uncertainty in the detected body position, the temporal prior is corrected at nodes newly exposed to the fluid by the moving body, and a coverage-adaptive smoothness term stabilizes the reconstruction in data-starved wake regions. When no body is present, all masking terms switch off and the formulation reduces to the underlying constrained reconstruction. LE-DM can also be used downstream of OA-LPT, which improves particle recovery at the tracking stage before the Eulerian reconstruction is performed.

LE-DM treats moving-boundary reconstruction as a single constrained problem, in which the geometry, boundary motion, data fidelity, incompressibility, temporal consistency, and smoothness are all combined in one step rather than handled separately. The signed-distance field is used not only as a geometric descriptor, but as the mechanism by which the reconstruction domain is updated in time and the roles of grid nodes are assigned as the body moves. Boundary kinematics are imposed as hard constraints on the boundary shell and solid interior, while incompressibility is restricted to the fluid through one-sided operators that prevent the constraint from crossing the moving surface. The data-fidelity, divergence, and smoothness operators are assembled on open-fluid nodes alone, so the constrained boundary values fix the discrete surface kinematics without entering the equations that determine the fluid field. This construction extends the constrained cost-minimization structure of \citet{agarwal2021reconstructing} to a time-dependent fluid-solid domain, while drawing on the sharp-interface logic of immersed-boundary methods \citep{mittal2005immersed,mittal2008versatile} and the use of hard no-slip constraints in fixed-boundary PTV reconstruction \citep{sperotto2022meshless,sperotto2024spicy}. The resulting formulation is not a post-processing mask applied to an existing reconstruction, but a single grid-based framework in which the moving body is part of the reconstruction problem itself. It handles translating bodies on a fixed Cartesian grid by updating the solid mask, reclassifying fluid and solid nodes, and correcting the temporal prior where the moving body exposes new fluid regions. This provides a complete reconstruction framework for moving boundaries without requiring a body-fitted mesh, a projection step, a separate boundary solver, or a post-solve correction.

The generality of LE-DM follows from the way the boundary is specified. The reconstruction requires only the instantaneous surface geometry and the corresponding surface velocity, which define the fluid-solid partition and the kinematic constraint. Fixed walls, translating or rotating bodies, and multiple bodies can therefore be treated within the same formulation without changing the reconstruction algorithm. In this study, the method is assessed using a sphere because this is the geometry for which high-quality volumetric measurements are available. Two cases provide reference fields for direct error evaluation. An analytical oscillating-sphere case gives an exact unsteady Stokes solution, and a CFD rising-sphere case provides the full simulated velocity field in a three-dimensional wake. The third case applies LE-DM to tomographic-PTV measurements of a freely rising sphere, where no independent reference field is available, and is used to demonstrate the complete workflow on real data. Together, these cases test whether the method recovers the prescribed surface motion, enforces incompressibility in the fluid, and produces a boundary-consistent velocity field next to the moving body.

% ─────────────────────────────────────────────────────────
\section{Methodology}
\label{sec:methodology}
% ─────────────────────────────────────────────────────────

LE-DM treats the moving solid as an active part of the Eulerian reconstruction. At each time step, a signed-distance field updates the fluid-solid partition on the fixed Cartesian grid and determines where data fitting, incompressibility, and boundary constraints should be applied. The measured tracks are used only in the fluid region, the divergence-free condition is restricted to the fluid domain, and the velocity on and inside the body is tied to the prescribed body motion. In this way, the reconstructed flow, the moving-boundary classification, and the physical constraints are assembled in one constrained problem rather than assembled through separate masking or correction steps. If the body is absent, the masking terms vanish and the method reverts to the original constrained reconstruction.

\subsection{Problem definition and dynamic mask}
\label{sec:domain_masking}

At each discrete time $t_k$, the particle-tracking algorithm provides $N_t$ particles with measured positions $\boldsymbol{\xi}_i(t_k)$ and velocities $\mathbf{v}_i(t_k)$, where $i=1,\ldots,N_t$. The particle velocities are stacked into the data vector $\mathbf{d}\in\mathbb{R}^{3N_t}$. The reconstruction is performed on a fixed Cartesian grid with node locations $\mathbf{r}_j$, where $j=1,\ldots,N_g$, and the unknown Eulerian velocity values at these nodes are collected in $\mathbf{q}\in\mathbb{R}^{3N_g}$. We use $\boldsymbol{\xi}_i$ exclusively for particle positions, $\mathbf{r}_j$ for grid-node positions, and $\mathbf{q}$ for the grid-based velocity unknown.

The instantaneous solid geometry is represented by a signed-distance field,
\begin{equation}
  \phi(\mathbf{r},t_k),
\end{equation}
following the standard level-set description of interfaces \citep{sethian1999level}. The sign convention is \(\phi<0\) inside the solid, \(\phi=0\) on the surface, and \(\phi>0\) in the fluid. This representation is also commonly used in sharp-interface immersed-boundary formulations, where a fixed Cartesian grid is retained while the body position is introduced through local geometric classification and boundary constraints \citep{mittal2005immersed,mittal2008versatile}. For a sphere of radius \(R\) centered at \(\mathbf{X}_s(t_k)\), \(\phi(\mathbf{r},t_k)=\|\mathbf{r}-\mathbf{X}_s(t_k)\|-R\). For an arbitrary body, \(\phi\) can be obtained from closest-point queries on a triangulated surface mesh. Multiple bodies are handled by taking the Boolean union of the individual solid regions, equivalently
\begin{equation}
  \phi(\mathbf{r},t_k)=\min_{m=1,\ldots,M}\phi_m(\mathbf{r},t_k).
\end{equation}

Each grid node is assigned to one of three disjoint sets,
\begin{equation}
\mathcal{C}_j(t_k)=
\begin{cases}
-1, & \phi(\mathbf{r}_j,t_k)<0, \qquad \text{solid interior},\\[3pt]
0,  & 0\leq \phi(\mathbf{r}_j,t_k)\leq \Delta/2, \qquad \text{boundary shell},\\[3pt]
+1, & \phi(\mathbf{r}_j,t_k)>\Delta/2, \qquad \text{open fluid},
\end{cases}
\label{eq:ternary}
\end{equation}
where $\Delta$ is the grid spacing. The corresponding node sets are denoted by
\begin{equation}
  \mathcal{F}(t_k)=\{j:\mathcal{C}_j=+1\},\quad
  \mathcal{S}(t_k)=\{j:\mathcal{C}_j=0\},\quad
  \mathcal{I}(t_k)=\{j:\mathcal{C}_j=-1\}.
\end{equation}
The shell is the fluid-side band of nodes lying within half a grid spacing of the surface, $0\le\phi_j\le\Delta/2$, rather than a literal one-cell-thick layer on the surface itself. The choice of $\Delta/2$ follows from the node-centered Cartesian discretization. A surface located between two grid nodes can be represented by the nearest fluid-side node only to within half a grid spacing, so this band identifies the nodes closest to the interface without extending the imposed boundary condition into the next layer of fluid. The kinematic boundary condition is imposed on this band, and the discrete interface is represented to within $\Delta/2$ of the true surface. This is the resolution limit associated with imposing boundary conditions on a fixed Cartesian grid.
 A fourth set is useful for assembling the divergence operator,
\begin{equation}
  \mathcal{N}(t_k)=\{j\in\mathcal{F}(t_k):\exists n\in\mathcal{A}(j)\;\text{with}\; n\notin\mathcal{F}(t_k)\},
\end{equation}
where $\mathcal{A}(j)$ denotes the axis-aligned nearest neighbors of node $j$. Nodes in $\mathcal{N}$ are open-fluid nodes whose centered finite-difference stencil would otherwise sample the shell or solid interior.

The prescribed solid velocity field is denoted by $\mathbf{u}_\Gamma(\mathbf{r},t_k)$. For a rigid body with translational velocity $\mathbf{U}_s(t_k)$ and angular velocity $\boldsymbol{\omega}_s(t_k)$,
\begin{equation}
  \mathbf{u}_\Gamma(\mathbf{r},t_k)
  =\mathbf{U}_s(t_k)+\boldsymbol{\omega}_s(t_k)\times
  \left[\mathbf{r}-\mathbf{X}_s(t_k)\right].
  \label{eq:u_gamma}
\end{equation}
For a stationary wall, $\mathbf{u}_\Gamma\equiv\mathbf{0}$. For multiple bodies, $m$, the velocity assigned near the interface is that of the closest body,
\begin{equation}
  \mathbf{u}_\Gamma(\mathbf{r},t_k)
  =\mathbf{u}_\Gamma^{(m^*)}(\mathbf{r},t_k),
  \qquad
  m^*=\arg\min_m \phi_m(\mathbf{r},t_k).
  \label{eq:multibody}
\end{equation}
This definition is adequate for separated bodies. If two interfaces approach within one grid spacing, the shell classification becomes under-resolved and the grid must be refined or the bodies treated as a merged solid for that time step.

\subsection{Constrained LE-DM functional}
\label{sec:system}

The Eulerian velocity field is recovered from the particle data through a constrained weighted least-squares problem. The formulation follows the constrained cost-minimization framework of \citet{agarwal2021reconstructing}, but the operators are assembled on the time-dependent masked domain introduced above. Thus, at each time $t_k$, the current fluid-solid classification determines where the data-fidelity, incompressibility, boundary-kinematic, and regularization terms are applied. LE-DM minimizes
\begin{equation}
\begin{split}
  \mathcal{J}(\mathbf{q},\boldsymbol{\lambda};t_k)
  =&\;[\mathbf{d}-\mathbf{A}(t_k)\mathbf{q}]^{\!\top}
       \mathbf{W}(t_k)
       [\mathbf{d}-\mathbf{A}(t_k)\mathbf{q}] \\
   &+\kappa[\mathbf{q}-\mathbf{q}_0(t_k)]^{\!\top}
       \mathbf{S}(t_k)
       [\mathbf{q}-\mathbf{q}_0(t_k)] \\
   &+\lambda_c\,\mathbf{q}^{\!\top}\mathbf{L}(t_k)\mathbf{q}
    +2\boldsymbol{\lambda}^{\!\top}
       [\mathbf{B}(t_k)^{\!\top}\mathbf{q}-\mathbf{g}(t_k)].
\end{split}
\label{eq:ccm_original}
\end{equation}
Here $\mathbf{A}$ interpolates the grid velocity to the measured particle locations, and $\mathbf{W}$ assigns the corresponding measurement weights. The vector $\mathbf{q}_0$ is the temporal reference field, $\mathbf{S}$ weights this temporal prior, and $\mathbf{L}$ provides spatial regularization on the masked grid. The matrix $\mathbf{B}^{\top}$ is the hard-constraint operator, with $\mathbf{g}$ containing the prescribed constraint values. The associated Lagrange multipliers $\boldsymbol{\lambda}$ enforce $\mathbf{B}^{\top}\mathbf{q}=\mathbf{g}$ exactly.

The first term measures the mismatch between the reconstructed grid velocity and the particle-track velocities. The second term introduces temporal regularization through the reference field $\mathbf{q}_0$, while the third provides spatial smoothing on the masked grid. These three terms determine the best-fit velocity field in regions where particle support may be sparse or noisy. The final term enforces the hard constraints through the Lagrange multipliers. In the open fluid, these constraints impose incompressibility. On the boundary shell and within the solid interior, they impose the prescribed rigid-body velocity. Hence, the solid interior is not treated as a fluid region. Its velocity is constrained only to remove inactive degrees of freedom and to prevent the reconstruction from assigning arbitrary values inside the body.

\subsection{Particle-to-grid mapping, measurement weighting, and regularization}
\label{sec:kernel}

The interpolation matrix $\mathbf{A}$ maps grid velocities to particle locations. The kernel choice is central to the robustness of the reconstruction under sparse sampling. A single-cell assignment, in which each particle influences only the eight nodes of the cell containing it, leaves empty cells with no direct data support, so the wake of a moving body can become discontinuous. A finite-support kernel instead allows each particle to contribute to a small neighborhood of nodes, so a locally empty cell can still receive information from nearby particles. For a particle at $\boldsymbol{\xi}_i$, the scalar interpolation weight associated with fluid grid node $j$ is \begin{equation} K_{ij}=\frac{\chi_j\,\psi\!\left[(\boldsymbol{\xi}_i-\mathbf{r}_j)/h\right]} {\displaystyle\sum_{n\in\mathcal{F}(t_k)} \chi_n\,\psi\!\left[(\boldsymbol{\xi}_i-\mathbf{r}_n)/h\right]}, \label{eq:kernel} \end{equation} where $\psi$ is a tensor-product cubic B-spline, $h$ is the kernel half-width, and $\chi_j=1$ for open-fluid nodes and zero otherwise. A cubic B-spline spans the nearest two grid cells in each direction, so each particle informs a compact neighborhood of nodes rather than a single cell. The same scalar weights are used independently for the three velocity components. The normalization gives a partition of unity over the available fluid support. Particles whose kernel has no open-fluid support are discarded at that time step. The support is clipped by the dynamic mask, $\chi_j$ removing all non-fluid nodes from both the numerator and the denominator, so particle information is never propagated across the solid interface. The diagonal entries of $\mathbf{W}$ combine the uncertainty of the particle velocity measurement with a surface-proximity damping factor. If $\sigma_{u,i}$ is the estimated uncertainty of the velocity of track $i$, and $\sigma_\Gamma(t_k)$ is the uncertainty in the detected body location, the weight assigned to each component of track $i$ is \begin{equation} W_i(t_k)=\frac{1}{\sigma_{u,i}^{2}} \left[1-\exp\!\left(-\frac{\max[0,\phi(\boldsymbol{\xi}_i,t_k)]}{\sigma_\Gamma(t_k)}\right)\right]. \label{eq:reweighting} \end{equation} Tracks far from the body retain their nominal inverse-variance weight. Tracks within one body-location uncertainty of the surface are down-weighted, while tracks reported on or inside the detected surface receive zero weight. For stationary boundaries, $\sigma_\Gamma$ is the calibration uncertainty of the wall location. If the boundary is analytically known, $\sigma_\Gamma$ may be set small relative to $\Delta$, in which case the weighting approaches a sharp mask. The temporal reference field $\mathbf{q}_0(t_k)$ is normally taken from the converged reconstruction at the previous time, $\mathbf{q}(t_{k-1})$. For a moving body, however, the classification of grid nodes changes continuously. To identify nodes that have just become non-solid, define \begin{equation} \tau_j(t_k)= \begin{cases} +1, & \mathcal{C}_j(t_k)\geq0\;\text{and}\;\mathcal{C}_j(t_{k-1})<0,\\[2pt] -1, & \mathcal{C}_j(t_k)<0\;\text{and}\;\mathcal{C}_j(t_{k-1})\geq0,\\[2pt] 0, & \text{otherwise}. \end{cases} \end{equation} At newly exposed nodes ($\tau_j=+1$), the previous solution corresponds to the velocity prescribed inside the solid and is therefore not a meaningful fluid prior. The opposite transition ($\tau_j=-1$), where a fluid node is overtaken by the body, requires no special prior because that node is reclassified as solid and is constrained directly to the body velocity. The regularization reference is reset only at newly exposed nodes, 
\begin{equation} \mathbf{q}_{0,j}(t_k)= \begin{cases} \mathbf{u}_\Gamma(\mathbf{r}_j,t_k), & \tau_j(t_k)=+1,\\[2pt] \mathbf{q}_{j}(t_{k-1}), & \text{otherwise}. \end{cases} 
\label{eq:prior_correction} \end{equation} 

This value is used only as a regularization reference. It is not a prescribed velocity at open-fluid nodes. For stationary geometries, $\tau_j=0$ everywhere and the correction is inactive. Regions that remain under-sampled over distances larger than the interpolation kernel require an additional spatial prior. This is introduced through a masked graph-Laplacian over open-fluid node pairs, \begin{equation} J_{\mathrm{smooth}}(t_k)=\lambda_c \sum_{(j,n)\in\mathcal{E}_F(t_k)} \bar{w}_{jn}(t_k)\, \left\|\mathbf{q}_j-\mathbf{q}_n\right\|^2, \label{eq:smooth} \end{equation} where $\mathcal{E}_F$ is the set of axis-aligned neighbor pairs for which both nodes are in $\mathcal{F}$, and $\bar{w}_{jn}=(w_j+w_n)/2$. The nodal weight decreases with local track support, \begin{equation} w_j(t_k)=\frac{1}{1+c_j(t_k)/c_0}, \end{equation} where $c_j$ is the number of tracks within one grid spacing of node $j$ and $c_0$ is a reference count. The term is weak in well-sampled regions and stronger where local track support is absent. In matrix form Eq.~\ref{eq:smooth} is $\lambda_c\mathbf{q}^{\top}\mathbf{L}\mathbf{q}$, where $\mathbf{L}$ is symmetric positive semi-definite and contains only fluid-fluid couplings. The value of $\lambda_c$ is selected once using a sparse synthetic benchmark with known ground truth and is then kept fixed for the experimental reconstructions.

\subsection{Interface constraints and saddle-point system}
\label{sec:kinematics}

The hard constraints are assembled from the node classification. At open-fluid nodes away from the interface, $j\in\mathcal{F}\setminus\mathcal{N}$, the corresponding row of $\mathbf{B}^{\top}$ contains the standard centered finite-difference approximation of
\begin{equation}
  \nabla\cdot\mathbf{u}=0.
\end{equation}
At shell nodes, the rigid-body kinematic condition is imposed as
\begin{equation}
  \mathbf{q}_j=\mathbf{u}_\Gamma(\mathbf{r}_j,t_k),\qquad j\in\mathcal{S}(t_k),
  \label{eq:noslip_hard}
\end{equation}
which is the discrete no-slip condition on the masked interface. At solid-interior nodes,
\begin{equation}
  \mathbf{q}_j=\mathbf{u}_\Gamma(\mathbf{r}_j,t_k),\qquad j\in\mathcal{I}(t_k),
  \label{eq:solid}
\end{equation}
This removes velocities in the non-fluid region from the reconstruction. Equations~\ref{eq:noslip_hard} and \ref{eq:solid} have the same algebraic form, but serve distinct roles. The shell rows impose the physical boundary condition, whereas the interior rows eliminate inactive solid degrees of freedom. The constrained interior nodes do not feed back into the fluid solution. The interpolation kernel is clipped by the mask, so no particle track contributes to an interior node. The smoothness Laplacian is assembled only over fluid-fluid neighbor pairs, and the near-interface divergence stencils are constructed to avoid sampling interior nodes. The interior constraints only fix otherwise unused degrees of freedom and prevent the interpolation and regularization terms from assigning arbitrary velocities inside the body volume. They do not influence the reconstructed fluid field.

For near-interface fluid nodes $j\in\mathcal{N}$, a centered divergence stencil may sample a shell or solid-interior node. The derivative in each coordinate direction is instead evaluated using only open-fluid nodes. This follows the numerical principle used in sharp-interface immersed-boundary methods, where the fluid equations are discretized on the fluid side of the interface and the boundary condition is imposed separately through the prescribed interface velocity \citep{mittal2005immersed,mittal2008versatile}. If the centered stencil in direction $x_\alpha$ crosses the solid and the side $s_{j\alpha}\in{-1,+1}$ has two open-fluid neighbors, a second-order one-sided derivative is used,

\begin{equation}
  \left.\frac{\partial u_\alpha}{\partial x_\alpha}\right|_j
  \approx
  \frac{-3u_{\alpha,j}
  +4u_{\alpha,j+s_{j\alpha}e_\alpha}
  -u_{\alpha,j+2s_{j\alpha}e_\alpha}}
  {2s_{j\alpha}\Delta_\alpha}.
  \label{eq:onesided}
\end{equation}
Directions for which both centered neighbors lie in the open fluid retain the centered-difference stencil. For highly concave or under-resolved geometries, the two open-fluid neighbors required by Eq.~\ref{eq:onesided} may not be available. In such cases, the implementation uses the highest-order admissible open-fluid stencil at that node. If no admissible stencil exists, the grid is flagged as insufficiently resolved. This prevents a fluid divergence row from coupling directly to a shell or solid-interior degree of freedom. The centered-divergence, one-sided-divergence, interpolation, and smoothness operators are assembled using only open-fluid nodes. The constrained shell and solid-interior values are then algebraically decoupled from the equations that determine the fluid field. The only case in which a divergence row could reach a boundary node is when a fluid node is adjacent both to the moving body and to the outer boundary of the reconstruction subvolume. In the reconstructions reported here, the subvolume is chosen large enough that the body remains fully inside it, so this configuration does not occur.

The constraint matrix and target vector are assembled by concatenating the divergence and kinematic blocks,
\begin{equation}
  \mathbf{B}(t_k)^{\!\top}=
  \begin{bmatrix}
    \mathbf{B}_{\mathrm{div}}^{F}(t_k)\\
    \mathbf{B}_{\mathrm{div}}^{N}(t_k)\\
    \mathbf{B}_{\mathrm{BC}}^{\mathrm{shell}}(t_k)\\
    \mathbf{B}_{\mathrm{BC}}^{\mathrm{solid}}(t_k)
  \end{bmatrix},
  \qquad
  \mathbf{g}(t_k)=
  \begin{bmatrix}
    \mathbf{0}\\
    \mathbf{0}\\
    \mathbf{g}_{\mathrm{BC}}^{\mathrm{shell}}(t_k)\\
    \mathbf{g}_{\mathrm{BC}}^{\mathrm{solid}}(t_k)
  \end{bmatrix}.
  \label{eq:B_assembly}
\end{equation}
The first two blocks enforce incompressibility only on open-fluid nodes. The last two blocks are row-selection operators whose right-hand sides contain the prescribed velocities from Eqs.~\ref{eq:noslip_hard} and \ref{eq:solid}. Thus, no grid node is simultaneously treated as open fluid and solid. This separation prevents the reconstruction from imposing mass conservation inside the solid and prevents interpolation or smoothing from propagating particle information through the body. The divergence, interpolation, and smoothness operators are all assembled on open-fluid nodes only, so the constrained shell and solid-interior values do not appear in the equations that determine the open-fluid field. Hence the shell rows impose the prescribed body velocity on the discrete boundary band, while each open-fluid value is fixed by the particle data, the masked incompressibility constraint, the temporal prior, and the fluid-side regularization.

Minimizing Eq.~\ref{eq:ccm_original} with respect to $\mathbf{q}$ and enforcing the constraints gives
\begin{equation}
  \begin{bmatrix}
    \mathbf{H}(t_k) & \mathbf{B}(t_k)\\[2pt]
    \mathbf{B}(t_k)^{\!\top} & \mathbf{0}
  \end{bmatrix}
  \begin{bmatrix}
    \mathbf{q}\\[2pt]
    \boldsymbol{\lambda}
  \end{bmatrix}
  =
  \begin{bmatrix}
    \mathbf{r}(t_k)\\[2pt]
    \mathbf{g}(t_k)
  \end{bmatrix},
  \label{eq:saddle_point}
\end{equation}
where
\begin{equation}
  \mathbf{H}=\mathbf{A}^{\top}\mathbf{W}\mathbf{A}
  +\kappa\mathbf{S}+\lambda_c\mathbf{L},
  \qquad
  \mathbf{r}=\mathbf{A}^{\top}\mathbf{W}\mathbf{d}
  +\kappa\mathbf{S}\mathbf{q}_0.
\end{equation}

Here $\mathbf{H}$ collects the data-fidelity, temporal-prior, and spatial-regularization operators acting on the unknown velocity field, while $\mathbf{r}$ contains the corresponding known contributions from the measured particle velocities and the temporal reference field. The resulting system is sparse and symmetric indefinite. Its upper-left block is symmetric positive semi-definite when the component matrices are assembled as described above. The hard constraints are solved simultaneously with the data assimilation, so the output field satisfies the discrete divergence constraints at the open-fluid nodes and the prescribed shell velocity to the relative tolerance of the linear solver. Thus, the statement that the field is divergence-free refers specifically to the open-fluid nodes, to solver tolerance. It is not a claim of pointwise mass conservation across the interface, where the one-sided stencils evaluate divergence on the fluid side only. No post-solve velocity correction is applied.

\subsection{Implementation, active quantities, and limiting cases}
\label{sec:generalization}

At each time step, LE-DM updates only the quantities that depend on the current body configuration, boundary motion, or particle distribution. Table~\ref{tab:quantities} summarizes these inputs and operators, and indicates how each enters the reconstruction. For a stationary wall, $\phi$ is fixed and $\mathbf{u}_\Gamma=0$. For a moving body, the signed-distance field, boundary-velocity target, transition flags, measurement weights, and mask-dependent operators are recomputed at each $t_k$.

\begin{table}[htbp]
  \centering
  \caption{Inputs and operators used in the LE-DM reconstruction. The same formulation is used for stationary and moving boundaries; only the signed-distance field and prescribed boundary velocity change.}
  \small
  \renewcommand{\arraystretch}{1.4}
  \setlength{\extrarowheight}{4pt}
  \begin{tabular}{p{0.23\textwidth} p{0.30\textwidth} p{0.37\textwidth}}
    \hline
    \textbf{Quantity} & \textbf{Obtained from} & \textbf{Use in the reconstruction} \\[4pt]
    \hline
    \begin{minipage}[t]{0.23\textwidth}\raggedright
    Particle data \(\boldsymbol{\xi}_i\), \(\mathbf{v}_i\), \(\sigma_{u,i}\)
    \end{minipage}
    &
    \begin{minipage}[t]{0.30\textwidth}\raggedright
    STB, OA-LPT, or another Lagrangian tracking method
    \end{minipage}
    &
    \begin{minipage}[t]{0.37\textwidth}\raggedright
    Supplies particle positions, velocity measurements, and measurement weights for the data-fidelity term.
    \end{minipage}
    \\[8pt]
    \begin{minipage}[t]{0.23\textwidth}\raggedright
    Kernel width \(h\)
    \end{minipage}
    &
    \begin{minipage}[t]{0.30\textwidth}\raggedright
    Chosen from particle spacing and grid spacing
    \end{minipage}
    &
    \begin{minipage}[t]{0.37\textwidth}\raggedright
    Sets the neighborhood over which particle information contributes to the Eulerian grid.
    \end{minipage}
    \\[8pt]
    \begin{minipage}[t]{0.23\textwidth}\raggedright
    Signed-distance field \(\phi(\mathbf{r},t_k)\)
    \end{minipage}
    &
    \begin{minipage}[t]{0.30\textwidth}\raggedright
    Analytical geometry, wall calibration, or tracked surface mesh
    \end{minipage}
    &
    \begin{minipage}[t]{0.37\textwidth}\raggedright
    Classifies grid nodes as open fluid, boundary shell, or solid interior.
    \end{minipage}
    \\[8pt]
    \begin{minipage}[t]{0.23\textwidth}\raggedright
    Boundary velocity \(\mathbf{u}_{\Gamma}(\mathbf{r},t_k)\)
    \end{minipage}
    &
    \begin{minipage}[t]{0.30\textwidth}\raggedright
    Prescribed wall motion or tracked rigid-body kinematics
    \end{minipage}
    &
    \begin{minipage}[t]{0.37\textwidth}\raggedright
    Provides the target velocity for shell and solid-interior constraints.
    \end{minipage}
    \\[8pt]
    \begin{minipage}[t]{0.23\textwidth}\raggedright
    Boundary-position uncertainty \(\sigma_{\Gamma}(t_k)\)
    \end{minipage}
    &
    \begin{minipage}[t]{0.30\textwidth}\raggedright
    Body-detection uncertainty or wall-calibration uncertainty
    \end{minipage}
    &
    \begin{minipage}[t]{0.37\textwidth}\raggedright
    Sets the damping length for near-interface particle measurements.
    \end{minipage}
    \\[8pt]
    \begin{minipage}[t]{0.23\textwidth}\raggedright
    Transition flag \(\tau_j(t_k)\)
    \end{minipage}
    &
    \begin{minipage}[t]{0.30\textwidth}\raggedright
    Current and previous node classifications
    \end{minipage}
    &
    \begin{minipage}[t]{0.37\textwidth}\raggedright
    Identifies nodes newly exposed by a moving body and resets their temporal prior.
    \end{minipage}
    \\[8pt]
    \begin{minipage}[t]{0.23\textwidth}\raggedright
    Constraint matrix \(\mathbf{B}\) and target vector \(\mathbf{g}\)
    \end{minipage}
    &
    \begin{minipage}[t]{0.30\textwidth}\raggedright
    Node classification, finite-difference stencils, and boundary kinematics
    \end{minipage}
    &
    \begin{minipage}[t]{0.37\textwidth}\raggedright
    Enforces incompressibility in the open fluid and prescribed boundary motion at the shell.
    \end{minipage}
    \\[8pt]
    \begin{minipage}[t]{0.23\textwidth}\raggedright
    Masked Laplacian \(\mathbf{L}(t_k)\)
    \end{minipage}
    &
    \begin{minipage}[t]{0.30\textwidth}\raggedright
    Fluid-node neighbor graph and local track support
    \end{minipage}
    &
    \begin{minipage}[t]{0.37\textwidth}\raggedright
    Provides spatial regularization in cells with weak local particle support without coupling through the solid.
    \end{minipage}
    \\[8pt]
    \hline
  \end{tabular}
  \label{tab:quantities}
\end{table}

\begin{figure}[t]
  \centering
  \includegraphics[width=0.99\columnwidth]{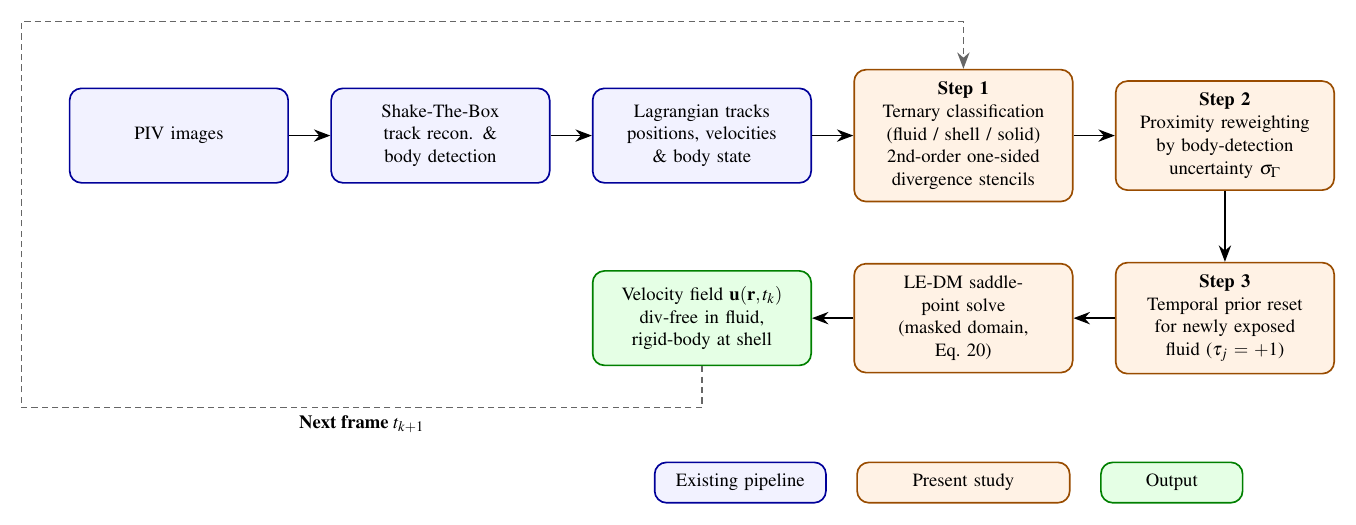}
  \caption{LE-DM processing pipeline for one time slab. The Lagrangian tracks and instantaneous body state provide the inputs. The signed-distance field defines the fluid, shell, and solid-interior node sets; the track distribution defines the interpolation, measurement weights, and coverage-adaptive smoothness; and the body kinematics define the shell and solid-interior targets. These quantities are assembled into the single saddle-point system of Eq.~\ref{eq:saddle_point}. The output is an Eulerian velocity field on the fixed grid, with discrete incompressibility imposed in the open fluid and rigid-body kinematics imposed at the boundary shell.}
  \label{fig:algorithm}
\end{figure}

At each time $t_k$, LE-DM starts from the particle tracks and the instantaneous body state provided by the tracking and body-detection pipeline. From the body state, the signed-distance field is evaluated, typically only in a bounding box surrounding the body or wall, following the standard level-set representation of immersed geometries \citep{sethian1999level}. This field is used to classify the grid into open-fluid, boundary-shell, and solid-interior nodes. The same classification then determines the near-wall nodes, the transition flags, the measurement weights, the corrected temporal prior, the masked Laplacian, and the constraint rows. Figure~\ref{fig:algorithm} summarizes this sequence for one time step.

The assembled system is the saddle-point problem in Eq.~\ref{eq:saddle_point}. It is solved with a Krylov method for symmetric indefinite systems, such as MINRES \citep{benzi2005numerical}, using the previous converged velocity field as a warm start after applying the prior correction. For stationary geometries, the mask and boundary-condition blocks do not change in time and can be assembled once. For moving bodies, these quantities are updated at every time step before the solve.

The formulation contains four user-specified parameters, each fixed by a stated rule rather than adjusted case by case. The kernel half-width $h$ is set by the cubic B-spline support, corresponding to two grid cells in each direction. The grid spacing $\Delta$ is chosen no smaller than the mean inter-particle spacing $\ell_p$, so that the reconstruction is controlled by the available tracer density rather than by an artificially fine grid, as discussed in Section~\ref{sec:val_oscsphere}. The Tikhonov weight $\kappa$ controls the strength of the temporal prior relative to the data term. It is held fixed within each dataset and chosen to stabilize poorly sampled regions without biasing well-sampled ones. The smoothness weight $\lambda_c$ is selected once using a sparse synthetic benchmark with known ground truth, at a tracer density representative of the experimental wake, and is then kept fixed for all reported reconstructions. The damping length $\sigma_\Gamma$ is not tuned. It is the per-snapshot body-position uncertainty supplied by the detection procedure. These rules ensure that the reported results do not rely on per-case parameter adjustment.

The computational cost is governed by the sparse saddle-point solve. The system has $3N_g+m(t_k)$ unknowns, where $N_g$ is the number of grid nodes and $m(t_k)$ is the number of active constraint rows. These constraints comprise one incompressibility row per open-fluid node that has a complete divergence stencil, with interior fluid nodes using a centered difference and near-wall fluid nodes using the one-sided stencil of Eq.~\ref{eq:onesided}, together with three kinematic rows per boundary-shell and solid-interior node. When the body moves, the constraint matrix is rebuilt at each time step because the mask changes. The matrix remains sparse, however, since each interpolation, divergence, and kinematic row involves only a local stencil of grid nodes.

Non-uniform track density can make the upper-left block poorly conditioned, because densely sampled regions carry much stronger data weights than empty or weakly sampled wake regions. The coverage-adaptive smoothness term improves the conditioning by adding spatial coupling where the data term provides little support, as discussed in Section~\ref{sec:val_piv}. With the previous converged field used as a warm start, the per-snapshot solve converges in a number of iterations small relative to the system size, since both the flow field and body position change gradually between consecutive frames.
For example, in the experimental case shown in Section \ref{sec:val_piv} using  $\Delta = 1$~mm, the reconstruction subvolume contains $N_g = 57{,}000$ nodes ($38\times50\times30$), of which $98.4\%$ are open fluid at a representative snapshot, with the remainder split between the boundary shell ($0.35\%$) and the solid interior ($1.3\%$). The constraint set comprises one divergence row per open-fluid node with a complete stencil and three kinematic rows per shell and solid-interior node, giving $m = 50{,}230$ rows and a saddle-point system of $3N_g + m = 221{,}230$ unknowns. The system matrix is sparse, with the upper-left block at $0.14\%$ density and a footprint of about $0.45$~GB in double precision. Starting from the previous converged frame and using no preconditioner, MINRES reaches the relative tolerance $\|\mathbf{r}\|/(\|\mathbf{K}\|\,\|\mathbf{x}\|)\le 10^{-6}$ in $80$ iterations, corresponding to a full-system relative residual of order $10^{-3}$ and a body-constraint residual below the $10^{-3}$ restart threshold. The single solve takes approximately $4.5$~s per snapshot on one core of an Intel i7-11700F processor.

The same equations cover several useful limiting cases. For a stationary wall, $\phi$ is fixed, $\mathbf{u}_\Gamma=0$, and the transition correction is inactive. For a translating body, $\boldsymbol{\omega}_s=0$ in Eq.~\ref{eq:u_gamma}. For a rotating or fully rigid-body motion, the complete form of Eq.~\ref{eq:u_gamma} is used. For multiple separated bodies, Eq.~\ref{eq:multibody} assigns each interface node to its closest body. For arbitrary shapes, only the evaluation of $\phi$ and $\mathbf{u}_\Gamma$ changes; the reconstruction system remains Eq.~\ref{eq:saddle_point}.

\section{Validation and demonstration}
\label{sec:validation}

The validation strategy is built around a sequence of spherical-body cases that increase in complexity from an exact analytical flow to synthetic particle tracks and finally to a tomographic-PTV experiment. This structure allows the main elements of LE-DM to be examined under controlled conditions before applying the method to the experimental configuration that motivated the work. The analytical case provides an exact reference field and isolates the moving-mask formulation from measurement uncertainty. The CFD-based case introduces an unsteady three-dimensional wake while retaining full access to the reference velocity field. The experimental case then tests the complete reconstruction workflow for a freely rising sphere in a refractive-index-matched solution, where finite seeding density, body-position uncertainty, and near-interface particle loss are all present.

The three cases are intended as a matched assessment of the moving-boundary formulation, rather than as a broad benchmark against all available reconstruction methods. The analytical oscillating sphere is used to verify the mask implementation in a setting where the exact solution is known. The CFD rising-sphere case is used to quantify reconstruction errors in the bulk flow and near the body as the tracer density is varied. The experimental rising-sphere case has no independent reference field and is evaluated through internal consistency diagnostics and comparison with the base CCM reconstruction obtained by disabling the mask. Quantitative errors are therefore reported only for the analytical and CFD cases. Because disabling the mask reduces LE-DM to base CCM, this comparison isolates the effect of the moving-boundary treatment, which acts at and near the interface and leaves the reconstruction unchanged in fully fluid regions. Away from the interface, the two reconstructions are expected to be similar. The relevant differences are those that occur near the moving boundary.

All three cases use a sphere because this is the geometry available in the high-quality volumetric experiment. Refractive-index matching is essential for this configuration because it is what makes volumetric particle tracking possible around a moving solid, by suppressing optical obstruction, path distortion, and the loss of near-surface particle information. The use of a sphere does not restrict the formulation itself. The body enters LE-DM through the signed-distance field and the prescribed kinematics, so other shapes can be treated by changing these inputs while leaving the reconstruction system unchanged. In this sense, the sphere serves as a controlled demonstration geometry rather than a limitation of the method.

Table~\ref{tab:validation_strategy} summarizes the role of the three cases. The analytical and CFD cases provide reference fields for quantitative error analysis, while the experimental case demonstrates the complete workflow under realistic measurement conditions.

\begin{table}[htbp]
\centering
\caption{Role of the three assessment cases used to evaluate the LE-DM reconstruction.}
\label{tab:validation_strategy}
\small
\renewcommand{\arraystretch}{1.25}
\begin{tabular}{p{0.22\textwidth} p{0.18\textwidth} p{0.50\textwidth}}
\hline
\textbf{Case} & \textbf{Reference field} & \textbf{Assessment focus} \\
\hline
\raggedright Oscillating sphere &
\raggedright Analytical solution &
\raggedright Verification of the moving-mask implementation using a closed-form velocity field, with emphasis on boundary kinematics, near-interface reconstruction, grid spacing, and Stokes-layer resolution. \tabularnewline

\raggedright Rising sphere, synthetic tracks &
\raggedright Numerical solution &
\raggedright Quantitative error analysis in a three-dimensional unsteady wake using particle-sampled velocity data, with separate evaluation of bulk-flow and near-body reconstruction accuracy. \tabularnewline

\raggedright Rising sphere, experiment &
\raggedright No independent reference field &
\raggedright End-to-end demonstration of the RIM-PTV workflow for a freely moving sphere under realistic experimental limitations, including finite seeding density, near-surface particle loss, and body-detection uncertainty. \tabularnewline

\hline
\end{tabular}
\end{table}

\subsection{Analytical benchmark: oscillating sphere in Stokes flow}
\label{sec:val_oscsphere}

The analytical test case provides the most controlled assessment of the moving-boundary formulation. It considers the unsteady Stokes flow generated by a rigid sphere undergoing prescribed harmonic motion in an otherwise quiescent fluid. Because the reference field is known exactly, the reconstruction can be evaluated without the influence of image reconstruction, track linking, body detection, calibration, or model uncertainty. This case is used to examine the dynamic mask under well-defined near-wall conditions, rather than to reproduce the wake dynamics of a freely rising sphere. The prescribed oscillatory motion provides a moving no-slip boundary, an exact velocity field for error evaluation, and a controllable Stokes-layer thickness that sets the severity of the near-surface gradient.

A rigid sphere of radius \(a=5.555\)~mm oscillates harmonically along the \(y\) direction, with translational velocity
\begin{equation}
  \mathbf{U}_s(t)
  =
  U_0 \cos(\omega t)\,\mathbf{e}_y ,
  \label{eq:osc_sphere_velocity}
\end{equation}
where \(U_0=0.02\)~m/s. The corresponding sphere-center position is
\begin{equation}
  \mathbf{X}_s(t)
  =
  \mathbf{X}_{s,0}
  +
  \frac{U_0}{\omega}\sin(\omega t)\,\mathbf{e}_y .
  \label{eq:osc_sphere_position}
\end{equation}

The exterior velocity field is taken as the unsteady Stokes solution for a rigid sphere oscillating in an otherwise quiescent viscous fluid \citep{landau1987fluid,kim1991microhydrodynamics}. This provides a closed-form reference field outside the body, allowing the reconstructed velocity to be compared directly with the exact solution. The fluid velocity is written as the real part of a time-harmonic field,
\begin{equation}
  \mathbf{u}^{\mathrm{exact}}(\mathbf{r},t)
  =
  \mathrm{Re}\!\left\{e^{i\omega t}\,\mathbf{u}^{*}(\mathbf{r})\right\}.
  \label{eq:stokes_osc}
\end{equation}
The complex amplitude $\mathbf{u}^{*}$ is written in spherical coordinates $(r,\theta)$ measured from the oscillation axis. With the complex wavenumber
\begin{equation}
  k=\sqrt{\frac{i\omega}{\nu}}=\frac{1+i}{\delta_{\mathrm{BL}}},
  \label{eq:wavenumber}
\end{equation}
the radial and tangential amplitudes are
\begin{equation}
  u^{*}_{r}=\frac{2\cos\theta}{r^{2}}\,F(r),
  \qquad
  u^{*}_{\theta}=-\frac{\sin\theta}{r}\,F'(r),
  \label{eq:stokes_components}
\end{equation}
where the radial function and its derivative are
\begin{equation}
  F(r)=\frac{B}{r}+C\,e^{-k(r-a)}\!\left(1+\frac{1}{kr}\right),
  \qquad
  F'(r)=-\frac{B}{r^{2}}-C\,e^{-k(r-a)}\!\left(k+\frac{1}{r}+\frac{1}{kr^{2}}\right),
  \label{eq:stokes_radial}
\end{equation}
and the complex coefficients, fixed by no slip at $r=a$ and quiescence as $r\to\infty$, are
\begin{equation}
  B=\tfrac{1}{2}U_0 a^{3}+\frac{3U_0 a^{2}}{2k}+\frac{3U_0 a}{2k^{2}},
  \qquad
  C=-\frac{3U_0 a}{2k}.
  \label{eq:stokes_coeffs}
\end{equation}
The term in $B$ is the potential-dipole part and the term in $C$ is the viscous, transverse part that decays over the Stokes-layer thickness $\delta_{\mathrm{BL}}$. The amplitudes are evaluated directly in this hand-expanded form, without spherical Hankel functions, and reduce to the rigid-surface velocity at $r=a$ to machine precision. Here $\theta$ is measured from the oscillation axis, which is taken as $+y$ in the laboratory frame, so the field of Eqs.~\ref{eq:stokes_components}--\ref{eq:stokes_coeffs} is mapped to laboratory Cartesian components with $\mathbf{e}_y$ as the polar axis. The spatial structure is controlled by a single dimensionless group, the Womersley number,
\begin{equation}
  \mathrm{Wo}
  =
  a\sqrt{\frac{\omega}{\nu}},
  \label{eq:Wo}
\end{equation}
which sets the unsteady Stokes-layer thickness,
\begin{equation}
  \delta_{\mathrm{BL}}
  =
  \sqrt{\frac{2\nu}{\omega}}
  =
  a\sqrt{2}\,\mathrm{Wo}^{-1}.
  \label{eq:delta_BL}
\end{equation}
Varying \(\mathrm{Wo}\) changes the boundary-layer thickness relative to the Eulerian grid without changing the body geometry. This makes the case useful for testing how the reconstruction behaves as the near-wall gradient becomes more difficult to resolve.

\subsubsection{Synthetic track generation.}

At each selected Womersley number, \(N_p=5\times10^4\) tracer particles are seeded uniformly at random inside a cubic domain of side length \(36\)~mm centered on the mean sphere position. Particles are excluded from a \(0.3\)~mm standoff region around the sphere surface. The resulting mean inter-particle spacing is \(\ell_p=0.97\)~mm. The seeding domain fully covers the reconstruction grid, so that the baseline assessment is not affected by missing particle support at the outer boundaries.

Twenty equally spaced snapshots are generated over one oscillation period, \(T=2\pi/\omega\). Particle positions are advanced using second-order Runge--Kutta integration of Eq.~\ref{eq:stokes_osc}. At every snapshot, the particle velocity is evaluated directly from the analytical field at the particle position. Hence, the resulting input is free of image-reconstruction, track-linking, and velocity-estimation uncertainties. The body input supplied to LE-DM consists only of the sphere center position at each time step. The sphere velocity used in the boundary condition is then obtained internally by finite differencing the supplied trajectory, matching the procedure used for experimental body tracks.

\subsubsection{Parameter sweeps.}

Three one-parameter sweeps are used to separate the effects of boundary-layer thickness, body-position uncertainty, and grid resolution. In the first sweep, $\mathrm{Wo}\in\{2,3\}$ is varied while holding the grid spacing at $\Delta=2$~mm and the body-position uncertainty parameter at $\sigma_\Gamma=0.5$~mm. These two cases correspond to $\delta_{\mathrm{BL}}/\Delta=1.96$ and $1.31$, respectively, and test the reconstruction as the near-wall gradient becomes less well resolved. In the second sweep, $\sigma_\Gamma$ is varied over $\{0.1,0.25,1.0,2.0\}$~mm at $\mathrm{Wo}=2$. This isolates the effect of the proximity weighting in the clean-data limit, where the particle positions and velocities are known exactly. In the third sweep, the grid spacing is varied over $\Delta\in\{2.0,1.0,0.5\}$~mm at $\mathrm{Wo}=3$. This tests the sensitivity of the reconstruction to the Eulerian grid resolution relative to the mean inter-particle spacing.

\subsubsection{Error metrics.}

The reconstruction is evaluated only at fluid nodes, \(\phi_j>0\). The primary metric is the RMS velocity error,
\begin{equation}
  \varepsilon_{L_2}(t_k)
  =
  \left[
  \frac{1}{|\mathcal{F}|}
  \sum_{j\in\mathcal{F}}
  \left\|
  \mathbf{u}^{\mathrm{LE\mbox{-}DM}}_j(t_k)
  -
  \mathbf{u}^{\mathrm{exact}}_j(t_k)
  \right\|^2
  \right]^{1/2},
  \label{eq:eps_L2}
\end{equation}
where \(\mathcal{F}\) is the set of open-fluid grid nodes. Errors are reported in non-dimensional form as \(\varepsilon_{L_2}/U_0\). To separate near-wall and bulk behavior, the error is also evaluated over a wall band, \(0<\phi\le\delta_{\mathrm{BL}}\), and over a bulk region, \(\phi>\delta_{\mathrm{BL}}\). This split links the error measure to the length scale of the analytical solution. The near-wall band contains the imposed moving boundary and the strongest velocity gradients, while the bulk region measures how much of the reconstruction error persists away from the interface.

Directional accuracy is assessed using a near-body alignment metric,
\begin{equation}
  \left\langle \cos\theta \right\rangle_{\mathrm{band}}(t_k)
  =
  \left\langle
  \frac{
  \mathbf{u}^{\mathrm{LE\mbox{-}DM}}_j(t_k)
  \cdot
  \mathbf{u}^{\mathrm{exact}}_j(t_k)
  }{
  \left\|
  \mathbf{u}^{\mathrm{LE\mbox{-}DM}}_j(t_k)
  \right\|
  \left\|
  \mathbf{u}^{\mathrm{exact}}_j(t_k)
  \right\|
  }
  \right\rangle_{0<\phi_j<2.5~\mathrm{mm}},
  \label{eq:cos_align}
\end{equation}
where nodes for which either velocity magnitude is below $0.05U_0$ are excluded. This threshold avoids direction estimates in locations where the velocity vector is too small for the angle to be meaningful. The alignment metric complements the RMS error by testing whether the reconstructed near-body velocity has the correct sign, phase, and orientation. For field-level comparison, the reconstructed and analytical velocity magnitudes are sampled on a ring of radius $a+2\Delta$ in the meridional plane. This ring lies beyond the first grid layer adjacent to the staircased mask, but remains close to the sphere surface. It provides a compact comparison of the angular structure of the near-body field without relying on a single point-wise error value.

\begin{figure}[tbp]
  \centering
 \includegraphics[width=0.99\textwidth]{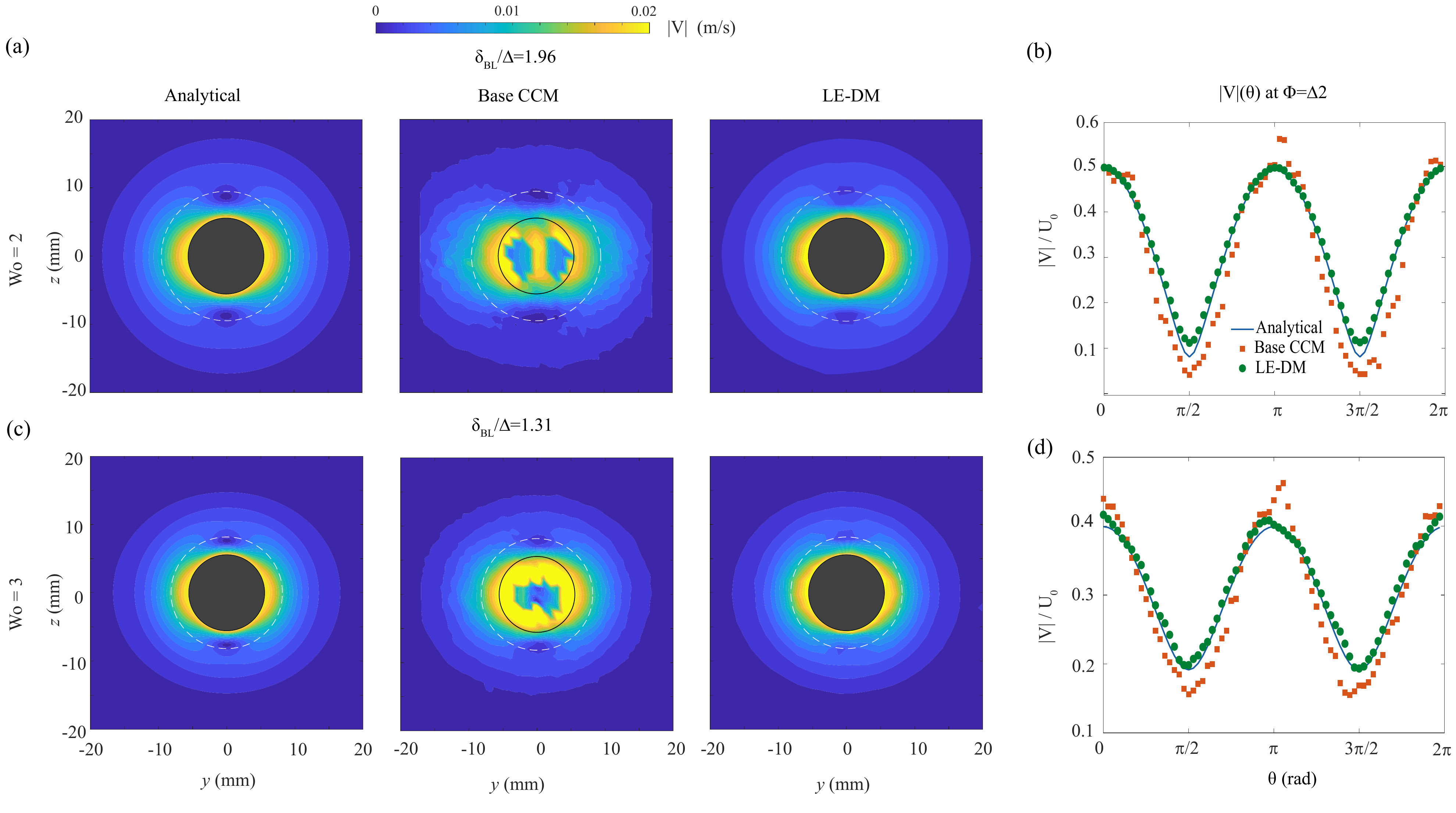}
  \caption{Velocity-field comparison on the meridional slice through the sphere center at the snapshot of peak sphere velocity, \(\omega t=0\). Top row: \(\mathrm{Wo}=2\); bottom row: \(\mathrm{Wo}=3\). Columns, left to right: analytical solution, base CCM reconstruction, LE-DM reconstruction, and velocity magnitude normalized by \(U_0\) versus polar angle \(\theta\) on a ring of radius \(a+2\Delta\). The angular profiles compare the analytical solution, the base CCM reconstruction, and LE-DM. The contour columns share a common color scale. The dashed circle marks the Stokes-layer thickness \(\delta_{\mathrm{BL}}\). In the analytical and LE-DM panels, the gray disc marks the sphere mask. In the base CCM reconstruction, the sphere interior is shown as reconstructed because no solid mask is supplied.}
  \label{fig:osc_field_comparison}
\end{figure}

\begin{figure}[htbp]
    \centering
    \includegraphics[width=.9\linewidth]{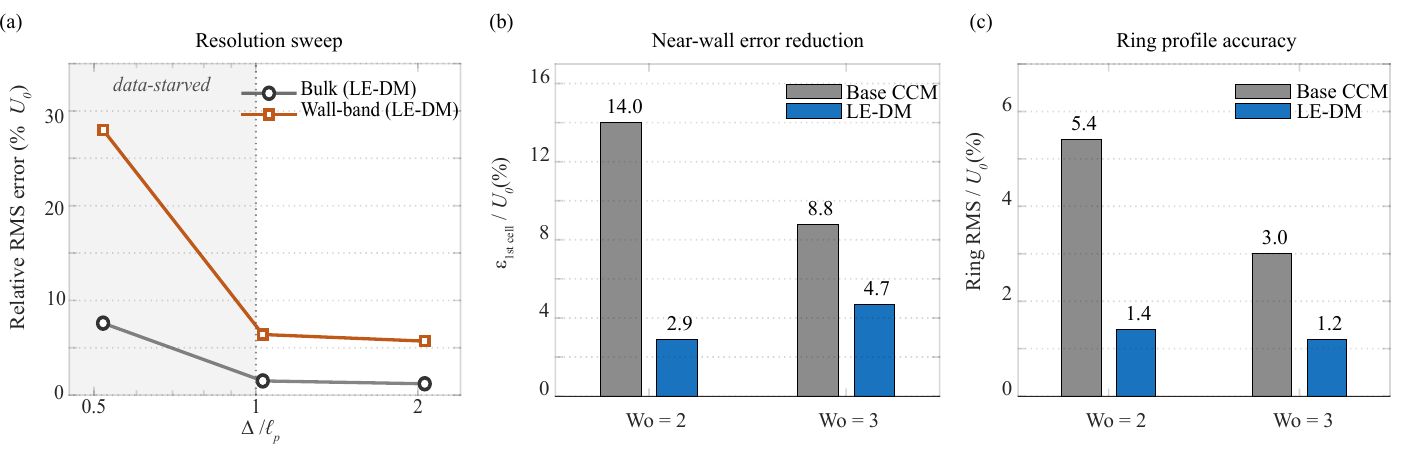}
    \caption{Resolution and near-interface sensitivity for the oscillating-sphere analytical assessment. \textbf{(a)} Relative RMS error versus \(\Delta/\ell_p\) for the bulk region and wall band at \(\mathrm{Wo}=3\); the dashed vertical line marks \(\Delta=\ell_p\). \textbf{(b)} First-cell error \(\varepsilon_{\mathrm{1st\,cell}}/U_0\) for the base CCM reconstruction and LE-DM at \(\mathrm{Wo}=2\) and \(\mathrm{Wo}=3\). \textbf{(c)} Ring-profile RMS deviation on the \(\phi=2\Delta\) sampling ring for the same two reconstructions. The comparison is diagnostic and is intended to show where the dynamic mask changes the near-interface solution.}
    \label{fig:osc_error_sweep}
\end{figure}

\subsubsection{Analytical reconstruction results.}

The analytical test confirms that LE-DM follows the prescribed oscillating-sphere field over the full cycle. The corresponding field comparison at peak sphere velocity is shown in Fig.~\ref{fig:osc_field_comparison}, where LE-DM preserves the masked body region and the near-body angular structure more closely than the base CCM reconstruction. The main error metrics are summarized in Table~\ref{tab:osc_validation}. For $\mathrm{Wo}=2$, the bulk error is $1.1\%$ of $U_0$, and the wall-band error is $2.9\%$. For $\mathrm{Wo}=3$, the bulk error decreases to $0.7\%$, while the wall-band error increases to $4.2\%$. This trend reflects the reduced Stokes-layer thickness at higher Womersley number. As the near-wall velocity gradient is confined to fewer grid cells, the wall band becomes more demanding to reconstruct, whereas a larger fraction of the domain remains in a slowly varying outer region. The resolution sweep in Fig.~\ref{fig:osc_error_sweep} shows the same behavior from a grid-resolution perspective. The bulk and wall-band errors remain low when the grid spacing is comparable to, or larger than, the mean inter-particle spacing. Both errors increase once the grid is refined below the available particle support. 

The alignment metric remains above $0.998$ throughout the cycle for both Womersley numbers. This shows that the near-body velocity direction is recovered accurately even when the wall-band magnitude error increases. The metric is especially useful near the zero crossing of the sphere velocity, where the imposed boundary velocity is small and the reconstruction is most sensitive to temporal regularization and possible sign errors in the boundary treatment.

The sweep in $\sigma_\Gamma$ produces little change in the bulk error for this clean analytical case. Across $\sigma_\Gamma=0.1$ to $2.0$~mm, the relative bulk error remains close to $1.75\%$. This weak dependence is expected because the particle positions, particle velocities, and body location are exact in this test. In this setting, the proximity weighting mainly affects the immediate interface neighborhood and does not control the bulk reconstruction. Its main role is expected in experimental data, where the near-interface particle field and the detected body surface carry finite measurement uncertainty.

\begin{table}[bp]
  \centering
  \caption{Oscillating-sphere analytical metrics. (using LE-DM)}
  \label{tab:osc_validation}
  \small
  \begin{tabular*}{\textwidth}{@{\extracolsep{\fill}}llccccc}
    \hline
    Sweep & Parameter & \(\delta_{\mathrm{BL}}/\Delta\) & \(\Delta/\ell_p\) & Bulk error & Wall error & Alignment \\
    \hline
    Womersley & \(\mathrm{Wo}=2\) & \(1.96\) & \(2.06\) & \(1.1\%\) & \(2.9\%\) & \(0.9997\) \\
    Womersley & \(\mathrm{Wo}=3\) & \(1.31\) & \(2.06\) & \(0.7\%\) & \(4.2\%\) & \(0.9988\) \\
    Damping & \(\sigma_\Gamma=0.1\)~mm & \(1.96\) & \(2.06\) & \(1.751\%\) & n/a & \(>0.999\) \\
    Damping & \(\sigma_\Gamma=0.25\)~mm & \(1.96\) & \(2.06\) & \(1.748\%\) & n/a & \(>0.999\) \\
    Damping & \(\sigma_\Gamma=1.0\)~mm & \(1.96\) & \(2.06\) & \(1.730\%\) & n/a & \(>0.999\) \\
    Damping & \(\sigma_\Gamma=2.0\)~mm & \(1.96\) & \(2.06\) & \(1.721\%\) & n/a & \(>0.999\) \\
    Resolution & \(\Delta=2.0\)~mm & \(1.31\) & \(2.06\) & \(1.2\%\) & \(5.7\%\) & \(0.999\) \\
    Resolution & \(\Delta=1.0\)~mm & \(2.62\) & \(1.03\) & \(1.5\%\) & \(6.4\%\) & \(0.999\) \\
    Resolution & \(\Delta=0.5\)~mm & \(5.24\) & \(0.52\) & \(7.6\%\) & \(28\%\) & \(0.86\) \\
    \hline
  \end{tabular*}

  \vspace{2mm}
  \footnotesize Bulk error is evaluated over \(\phi>\delta_{\mathrm{BL}}\), and wall error over \(0<\phi\le\delta_{\mathrm{BL}}\). Both are normalized by \(U_0\). Alignment denotes the median value of \(\left\langle \cos\theta \right\rangle_{\mathrm{band}}\) over the oscillation cycle. All cases use \(a=5.555\)~mm, \(U_0=0.02\)~m/s, \(\nu=10^{-5}\)~m\(^{2}\)/s, and \(\ell_p=0.97\)~mm.
\end{table}

The grid-spacing sweep shows that the reconstruction accuracy is limited by particle presence, not by grid spacing alone. Reducing $\Delta$ from $2.0$~mm to $1.0$~mm leaves the bulk error nearly unchanged, with values between $1.1\%$ and $1.5\%$ of $U_0$. In this range, the grid remains supported by the available particle distribution. Further refinement to $\Delta=0.5$~mm gives $\Delta/\ell_p=0.52$, so many grid cells have little or no local particle support. The bulk error then increases to $7.6\%$, and the wall-band error increases to $28\%$. This result highlights a practical constraint for PTV-based reconstruction. Refining the Eulerian grid well below the particle spacing does not necessarily improve the recovered field unless the additional degrees of freedom are supported by more particle data, stronger regularization, or multi-frame information.

\subsubsection{Diagnostic comparison with the base CCM reconstruction.}

The base CCM reconstruction is used as a diagnostic reference to isolate the effect of the dynamic mask. It receives the same synthetic particle input as LE-DM, but no solid geometry or boundary kinematics are supplied. The comparison in Fig.~\ref{fig:osc_field_comparison} and Table~\ref{tab:ccm_vs_ledm} is intended to show where the additional moving-boundary information enters the solution. Away from the sphere, where the boundary condition is not active, the two reconstructions are expected to give similar results.

The differences are concentrated near the moving interface and within the region occupied by the sphere. In the base CCM reconstruction, the sphere interior remains part of the reconstruction domain, so a velocity field is inferred inside the body even though that region contains no fluid. The near-surface field is then determined only by the surrounding particle data and by regularization. In LE-DM, the mask removes the solid interior from the fluid reconstruction and constrains the boundary shell to the prescribed rigid-body velocity. This supplies the missing geometric and kinematic information at the interface, allowing the exterior near-body field to follow the analytical solution more closely while leaving the bulk field largely unchanged.

At the snapshot of peak sphere velocity, LE-DM reduces both the first-cell error and the ring-profile deviation relative to the base CCM reconstruction, as shown in Fig.~\ref{fig:osc_error_sweep}b,c. For $\mathrm{Wo}=2$, the first-cell error decreases from $14.0\%$ to $2.9\%$ of $U_0$, and the ring-profile RMS deviation decreases from $5.4\%$ to $1.4\%$ of $U_0$. For $\mathrm{Wo}=3$, where the Stokes layer is thinner, the corresponding ring-profile RMS deviation decreases from $3.0\%$ to $1.2\%$ of $U_0$. These diagnostics emphasize the near-interface role of the dynamic mask. In this clean benchmark, the base CCM reconstruction already performs well in the bulk because the particle field is dense and noise-free. The gain from LE-DM is localized primarily to the moving boundary, where the prescribed mask and rigid-body velocity remove the non-physical interior reconstruction and improve the exterior near-wall field.

\begin{table}[b]
  \centering
  \caption{Diagnostic comparison between the base CCM reconstruction and LE-DM for the oscillating-sphere case.}
  \label{tab:ccm_vs_ledm}
  \small
  \begin{tabular*}{\textwidth}{@{\extracolsep{\fill}}lcccc}
    \hline
    Metric & base CCM, \(\mathrm{Wo}=2\) & LE-DM, \(\mathrm{Wo}=2\) & base CCM, \(\mathrm{Wo}=3\) & LE-DM, \(\mathrm{Wo}=3\) \\
    \hline
    Bulk error & \(3.0\%\) & \(1.1\%\) & \(0.9\%\) & \(0.7\%\) \\
    First-cell error & \(14.0\%\) & \(2.9\%\) & \(8.8\%\) & \(4.7\%\) \\
    Ring RMS & \(5.4\%\) & \(1.4\%\) & \(3.0\%\) & \(1.2\%\) \\
    Peak deviation & \(+2.3\%\) & \(+1.5\%\) & \(+9.6\%\) & \(+0.5\%\) \\
    \hline
  \end{tabular*}

  \vspace{2mm}
  \footnotesize Both reconstructions use the same synthetic particle data and are evaluated at the snapshot of peak sphere velocity. Bulk error is evaluated over \(\phi>\delta_{\mathrm{BL}}\), first-cell error over the grid cell adjacent to the surface, ring RMS on the \(\phi=2\Delta\) sampling ring, and peak deviation at the velocity maximum. All errors are normalized by \(U_0\).
\end{table}

The analytical case provides a controlled baseline for evaluating the moving-boundary components of LE-DM. It shows, first, that the imposed boundary motion is recovered with the correct direction and phase, with the near-body alignment metric remaining above $0.998$ throughout the oscillation cycle. Second, the error trends follow the expected dependence on Stokes-layer thickness. The bulk error remains small, while the near-wall error increases when the velocity gradient is compressed into fewer grid cells. Third, the grid-refinement study shows that the useful Eulerian resolution is limited by the available particle support. Refining the grid below the mean inter-particle spacing creates a data-starved reconstruction rather than a more accurate one. These observations establish the expected behavior of the dynamic mask under idealized conditions before realistic particle sampling, body-detection uncertainty, and experimental optical limitations are introduced in the following cases.

% ---------------------------------------------------------------------
\subsection{Synthetic Lagrangian tracks from CFD: turbulent rising sphere}
\label{sec:val_cfd}
% ---------------------------------------------------------------------

This case introduces an unsteady three-dimensional wake and particle-sampling error while retaining access to a known reference field. Synthetic Lagrangian tracks are generated from a CFD simulation of a buoyant rising sphere. The simulated configuration is chosen to be dynamically comparable to the rising-sphere experiment described in Section~\ref{sec:val_piv}. The sphere diameter, density ratio, viscosity, and body forcing are selected to place the simulation in the same high-Reynolds-number rising-sphere regime as the experiment. Since the complete numerical flow field is available, the reconstruction error can be evaluated directly and decomposed into bulk and near-body contributions. This provides a turbulent-flow validation that is not available from the analytical test case alone.

\subsubsection{CFD simulation parameters.}
The reference flow field is generated with an in-house three-dimensional lattice Boltzmann method (LBM) solver for a freely rising sphere. The solver uses a D3Q19 lattice with a regularized BGK collision model, following the regularized LBM formulation of \citet{cfd-latt2006lattice}. The computational domain contains $150 \times 500 \times 150$ lattice nodes in the horizontal, vertical, and spanwise directions, respectively. The sphere radius is $R=6$ lattice units, corresponding to a diameter $D=12$ lattice units. The sphere is resolved by 12 lattice nodes per diameter, and the domain size is approximately $12.5D \times 41.7D \times 12.5D$. The sphere is initially positioned at the center of the horizontal plane and near the lower part of the domain, with initial center location
\[
(x_0,y_0,z_0) = (75,31,75)
\]
in lattice units.

The fluid density is set to $\rho_f=1$ in lattice units, and the solid-to-fluid density ratio is
\[
\rho_s/\rho_f = 0.65 .
\]
The kinematic viscosity is
\[
\nu = 2\times 10^{-4},
\]
and the relaxation frequency is computed from
\[
\omega =
\left(
\frac{\nu}{c_s^2}+\frac{1}{2}
\right)^{-1},
\qquad
c_s^2=\frac{1}{3}.
\]
Gravity is imposed using the forcing formulation of \citet{cfd-guo2002discrete}. The body force is defined from the prescribed Galileo number, density contrast, viscosity, and sphere diameter. The CFD reference case is run at
\[
Ga = 2000,
\]
with density ratio $\rho_s/\rho_f=0.65$, placing the simulation in an unsteady, path-unstable rising-sphere regime. In the lattice simulation, the instantaneous Reynolds number is monitored as
\[
Re(t)=\frac{|U_s(t)|D}{\nu},
\]
where $U_s(t)$ is the vertical velocity of the sphere. The use of Galileo number and density ratio as controlling parameters for freely rising or falling spheres follows standard rising-sphere studies \citep{cfd-veldhuis2004motion, horowitz2010effect}.

The moving spherical boundary is represented using a sharp immersed-boundary bounce-back treatment based on interpolated bounce-back ideas for curved and moving boundaries \citep{cfd-bouzidi2001momentum}. A signed-distance function $\phi$ is constructed about the instantaneous sphere center, with $\phi<0$ denoting the solid region and $\phi>0$ denoting the fluid region. The immersed-boundary link data are rebuilt whenever the sphere center has moved by more than 0.25 lattice units. The moving-wall velocity used in the boundary treatment is taken from the instantaneous rigid-body velocity of the sphere. The hydrodynamic force obtained from the immersed-boundary treatment is combined with the effective buoyancy force,
\[
\mathbf{F}_g=(\rho_s-\rho_f)V_s\mathbf{g},
\]
where $V_s=(4/3)\pi R^3$ is the sphere volume. The resulting total force is used to update the sphere velocity and position explicitly in time.

Open boundary conditions are imposed by copying the distribution functions from the adjacent interior planes. Sponge damping is also applied near the streamwise boundaries to relax the solution toward a quiescent reference state and reduce reflections from the finite computational domain. The simulation is advanced for 20,000 lattice time steps. Eulerian velocity fields, the signed-distance field, and the immersed-boundary volume-fraction field are exported every 200 time steps on the full structured Cartesian grid.

The LBM simulation is performed in lattice units. Conversion to physical units is obtained by matching the numerical sphere diameter and fluid kinematic viscosity to the corresponding experimental values. The physical lattice spacing is
\[
\Delta x = \frac{D_{\rm exp}}{D_{\rm LB}},
\qquad D_{\rm LB}=12,
\]
and the physical time step follows from viscosity scaling,
\[
\Delta t =
\nu_{\rm LB}
\frac{\Delta x^2}{\nu_{\rm exp}} .
\]
Positions, times, and velocities are then converted according to
\[
\mathbf{x}_{\rm phys}=\mathbf{x}_{\rm LB}\Delta x,
\qquad
t_{\rm phys}=t_{\rm LB}\Delta t,
\qquad
\mathbf{u}_{\rm phys}
=
\mathbf{u}_{\rm LB}
\frac{\Delta x}{\Delta t}.
\]

The exported Eulerian velocity fields provide the reference data for synthetic tracer generation and for the subsequent reconstruction-error analysis. The internally integrated sphere center from the CFD solver is used as the simulation reference trajectory for calibrating and evaluating the sphere-location reconstruction algorithm. In the present study, the CFD field is used as a controlled numerical reference: unlike the experiment, the full Eulerian velocity field and the simulated sphere trajectory are known at every exported time, allowing the reconstruction error to be measured directly.

\subsubsection{Synthetic dataset generation.}
For each particle-density case, initial particle positions are sampled uniformly throughout the test volume, and particles inside the sphere are removed. The particles are then advanced over steps of $\Delta t=10$~ms using fourth-order Runge--Kutta integration of the simulated velocity field, interpolated at the particle locations with modified Akima interpolation. Because the CFD field has limited resolution in the immediate near-surface layer, a small corrective displacement is applied to any particle that would otherwise penetrate the moving surface. This enforces the no-penetration condition without altering the surrounding wake. Particle velocities are then obtained from a fourth-order central finite difference of the trajectories. This procedure reproduces the two features of real STB output that are most relevant for the reconstruction: scattered, finite-density sampling and the absence of particles inside the body.

\subsubsection{Parameter sweep.}
A parameter sweep is performed over grid spacing $\Delta \in \{1.0,\, 1.5,\, 2.0\}$~mm and tracer count. Each resolution $\Delta$ is applied to particle datasets with $N_p\in \{0.3, 0.5, 1.0, 2.0, 5.0\}\times 10^6$, resulting in mean inter-particle spacings of $\ell_p = \{2.6, 1.6, 0.78, 0.39, 0.16\}$~mm, respectively. The remaining LE-DM parameters are fixed at $\sigma_\Gamma=0.01$~mm, $\sigma_{u,i}=0.005$~m/s, and $\kappa=10$.

\subsubsection{Error metrics.}
The reconstructed velocity field is compared with the simulation output after interpolation onto the coarser grid using modified Akima interpolation. Relative RMS error, normalized by the sphere velocity $\varepsilon_{L^2}/U_0$, and the mean deviation angle $\theta$ between the simulation and reconstruction are averaged over the final five snapshots. The deviation angle is computed from the vector-alignment metric $\langle\cos{\theta}\rangle$ defined in Eq.~\ref{eq:cos_align} and is reported in degrees. The test volume is also divided into two sub-volumes, the wake region and the wall region, and the same error metrics are evaluated in both domains. In each snapshot, the wake region is defined as the minimal cuboid containing all grid nodes with a fluid speed of at least 15\% of $U_0$. The remaining cells outside the wake region are classified as the wall region.

\begin{figure}
    \centering
    \includegraphics[width=.95\linewidth]{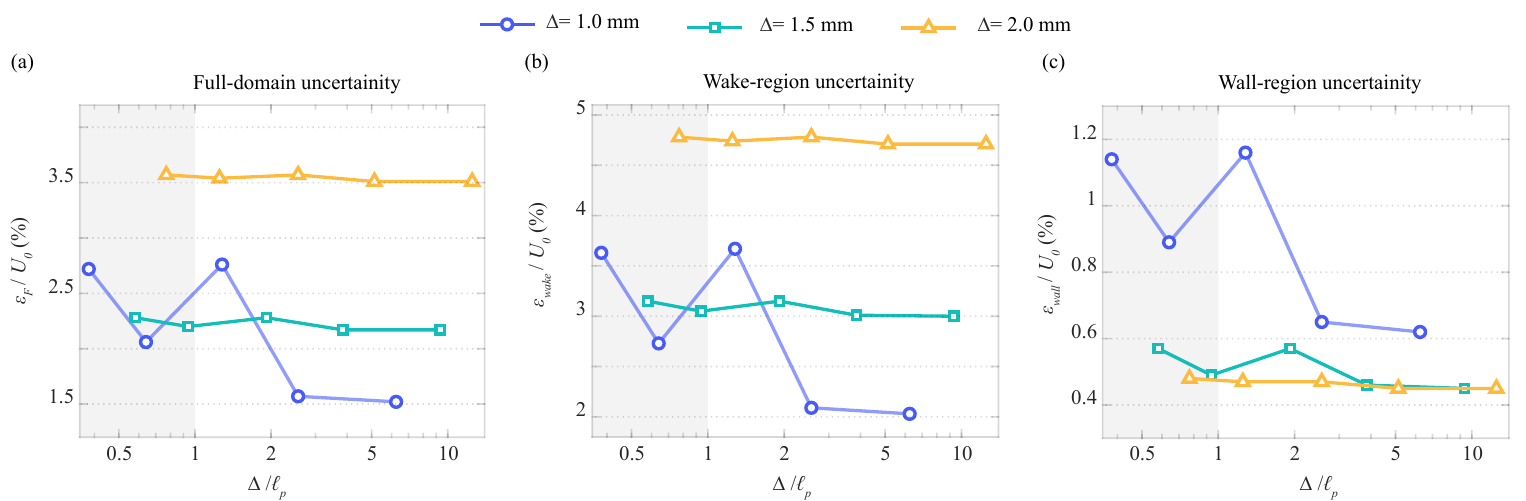}
    \caption{Reconstruction error versus tracer density for the CFD rising-sphere validation case.
             \textbf{(a)} Full-domain relative RMS error $\varepsilon_{\mathcal{F}}/U_0$,
             \textbf{(b)} wake-region error $\varepsilon_{\mathrm{wake}}/U_0$, and
             \textbf{(c)} wall-region error $\varepsilon_{\mathrm{wall}}/U_0$, each plotted against $\Delta/\ell_p$ for grid spacing $\Delta = 1.0$~mm (circles), $1.5$~mm (triangles), and $2.0$~mm (squares).
             Errors are averaged over the final five snapshots of each run.
             The wall-region error saturates at low $\Delta/\ell_p$, while the wake error continues to improve with increasing tracer density, reflecting the stronger velocity gradients concentrated in the near-wake.}
    \label{fig:cfd_error_sweep}
\end{figure}

\begin{table}[ht]
  \centering
  \caption{Validation metrics for reconstruction of the CFD simulated field.}
  \label{tab:cfd_validation}
  \begin{tabular}{@{}lccccccccc@{}}
    \toprule
    Resolution\!& $N_p$ & $\rho_p$ & $\Delta/\ell_p$ & $\varepsilon_{\mathcal{F}}/U_0$ & $\varepsilon_{\mathrm{wake}}/U_0$ & $\varepsilon_{\mathrm{wall}}/U_0$ & $\theta_{\mathcal{F}}$ & $\theta_{\mathrm{wake}}$ & $\theta_{\mathrm{wall}}$ \\
    \midrule
    & $3\times10^5$ & $0.336$ & $0.38$ & $2.72\%$ & $3.63\%$ & $1.14\%$ & $5.06^\circ$ & $5.62^\circ$ & $4.51^\circ$ \\
    $\Delta = 1.0$~mm\!\!& $5\times10^5$ & $0.560$ & $0.64$ & $2.06\%$ & $2.73\%$ & $0.89\%$ & $3.62^\circ$ & $6.44^\circ$ & $3.89^\circ$ \\
    & $1\times10^6$ & $1.12$ & $1.28$ & $2.76\%$ & $3.67\%$ & $1.16\%$ & $5.19^\circ$ & $5.67^\circ$ & $4.66^\circ$ \\
    I=$50$, J=$255$,\!\!& $2\times10^6$ & $2.24$ & $2.56$ & $1.57\%$ & $2.09\%$ & $0.65\%$ & $2.92^\circ$ & $2.14^\circ$ & $3.53^\circ$ \\
    K=$70$& $5\times10^6$ & $5.60$ & $6.25$ & $1.52\%$ & $2.03\%$ & $0.62\%$ & $2.81^\circ$ & $2.14^\circ$ & $3.53^\circ$\\
    \midrule
    & $3\times10^5$ & $1.14$ & $0.58$ & $2.28\%$ & $3.15\%$ & $0.57\%$ & $4.58^\circ$ & $5.56^\circ$ & $3.53^\circ$ \\
    $\Delta = 1.5$~mm\!\!& $5\times10^5$ & $1.90$ & $0.94$ & $2.20\%$ & $3.05\%$ & $0.49\%$ & $4.58^\circ$ & $5.50^\circ$ & $3.44^\circ$ \\
    & $1\times10^6$ & $3.79$ & $1.92$ & $2.28\%$ & $3.15\%$ & $0.57\%$ & $4.66^\circ$ & $5.56^\circ$ & $3.53^\circ$ \\
    I=$33$, J=$170$,\!\!& $2\times10^6$ & $7.59$ & $3.85$ & $2.17\%$ & $3.01\%$ & $0.46\%$ & $4.44^\circ$ & $5.44^\circ$ & $3.24^\circ$ \\
    K=$47$& $5\times10^6$ & $19.0$ & $9.34$ & $2.17\%$ & $3.00\%$ & $0.45\%$ & $2.81^\circ$ & $2.14^\circ$ & $3.53^\circ$ \\
    \midrule
    & $3\times10^5$ & $2.68$ & $0.77$ & $3.57\%$ & $4.78\%$ & $0.48\%$ & $6.78^\circ$ & $8.51^\circ$ & $3.71^\circ$ \\
    $\Delta = 2.0$~mm\!\!& $5\times10^5$ & $4.46$ & $1.25$ & $3.54\%$ & $4.74\%$ & $0.47\%$ & $6.73^\circ$ & $8.47^\circ$ & $3.62^\circ$ \\
    & $1\times10^6$ & $8.93$ & $2.56$ & $3.57\%$ & $4.78\%$ & $0.47\%$ & $6.73^\circ$ & $8.47^\circ$ & $3.71^\circ$ \\
    I=$25$, J=$128$,\!\!& $2\times10^6$ & $17.9$ & $5.13$ & $3.51\%$ & $4.71\%$ & $0.45\%$ & $6.64^\circ$ & $8.43^\circ$ & $3.53^\circ$ \\
    K=$25$& $5\times10^6$ & $44.6$ & $12.5$ & $3.51\%$ & $4.71\%$ & $0.45\%$ & $6.64^\circ$ & $8.43^\circ$ & $3.53^\circ$ \\
    \bottomrule
  \end{tabular}
\end{table}

\subsubsection{Results.}

The full sweep is summarized in Table~\ref{tab:cfd_validation} and Fig.~\ref{fig:cfd_error_sweep}. The lowest full-domain errors are $1.52\%$, $2.17\%$, and $3.51\%$ of $U_0$ for $\Delta=1.0$, $1.5$, and $2.0$~mm, respectively. The reconstruction accuracy improves as the grid spacing approaches the available tracer spacing and as the tracer density increases. The error is spatially localized rather than uniformly distributed across the domain. It is largest in the near wake and close to the sphere, where the velocity gradients are steepest, while the reconstructed field remains well aligned with the reference solution in the outer region. This indicates that the particle sampling is sufficient to recover the large-scale wake structure even at modest tracer density. Increasing the tracer count reduces the wake error most strongly. By contrast, the wall-region error tends to saturate once the inter-particle spacing falls below the grid spacing, because additional particles provide limited independent support within each grid cell. This case primarily tests LE-DM in the unsteady wake and near-body regions, where the absence of a moving-boundary treatment has the largest effect, rather than in a smooth background flow.

\begin{figure}
    \centering
    \includegraphics[width=1\linewidth]{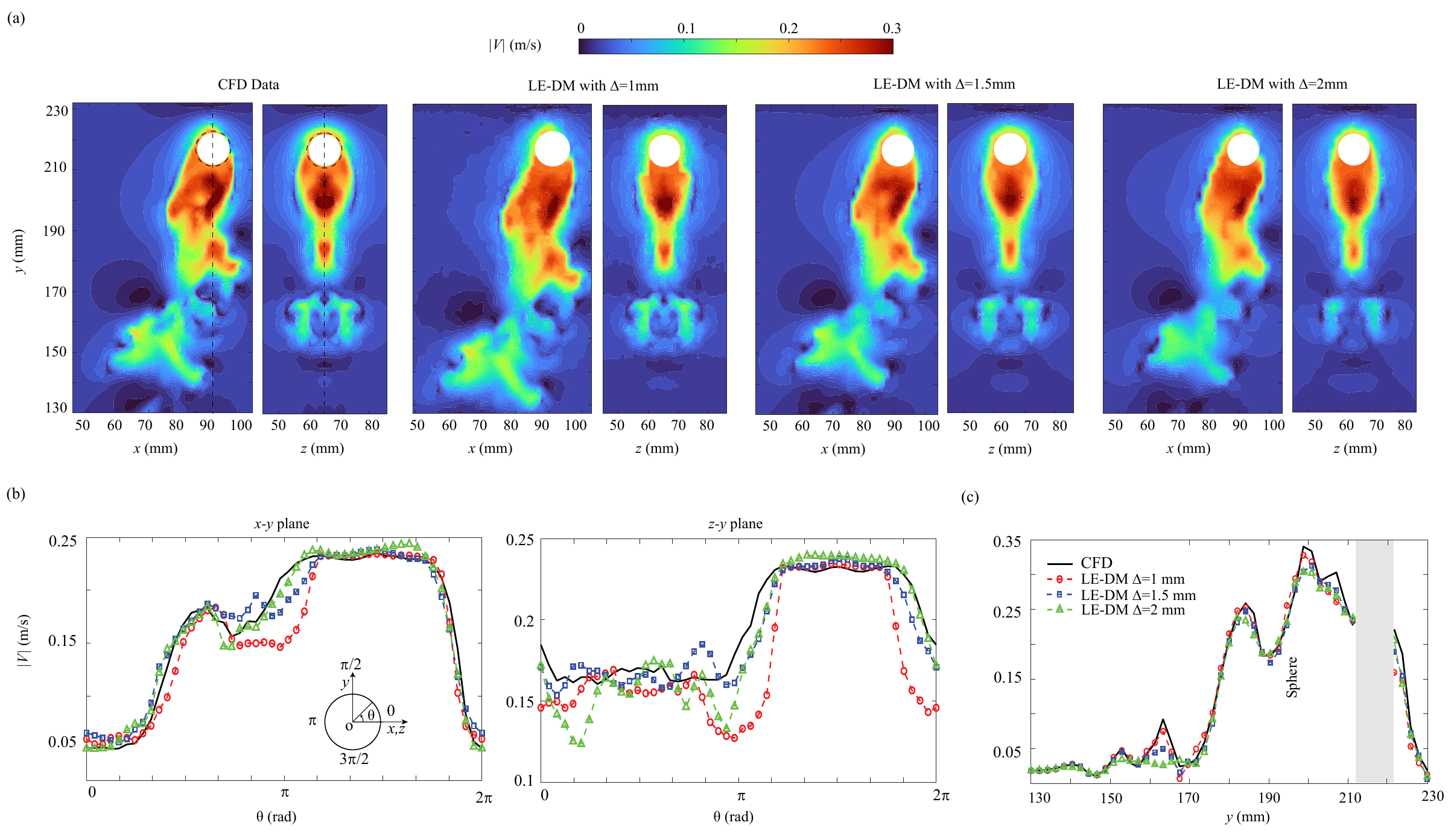}
    \caption{Reconstructed velocity field for the CFD synthetic-track validation case at the lowest tracer density, $N_p=3\times10^5$. \textbf{(a)} Velocity magnitude on the $x$--$y$ and $z$--$y$ meridional slices through the sphere center, comparing the CFD reference field with the LE-DM reconstruction at grid spacing $\Delta=1.0$, $1.5$, and $2.0$~mm on a common color scale. \textbf{(b)} Velocity magnitude versus polar angle $\theta$ sampled around the wake for the same two planes, comparing the CFD reference with LE-DM at the three grid spacings. \textbf{(c)} Velocity magnitude along the rise axis through the wake centerline. The reconstruction tracks the CFD reference closely through the bulk wake; residual differences concentrate in the steep-gradient region next to the sphere, where at the finest grid the $\Delta/2=0.5$~mm shell band approaches the resolution of the CFD reference itself, consistent with the resolution and tracer-density trends in Table~\ref{tab:cfd_validation}.}
    \label{fig:cfd_field_comparison}
\end{figure}

The spatial structure of the reconstruction is shown in Fig.~\ref{fig:cfd_field_comparison} for the lowest tracer density, $N_p=3\times10^5$, where the reconstruction is most strongly limited by particle support. Figure.~\ref{fig:cfd_field_comparison}a compares the CFD reference velocity magnitude with the LE-DM reconstruction at the three grid spacings on the $x$--$y$ and $z$--$y$ meridional slices. At all three grid spacings, LE-DM recovers the high-speed wake column above the sphere and the asymmetric near-wake structure below it. Figure.~\ref{fig:cfd_field_comparison}b compares the velocity magnitude as a function of polar angle around the wake in the two meridional planes. Figure.~\ref{fig:cfd_field_comparison}c compares the velocity magnitude along the rise axis through the wake centerline. The reconstructed profiles follow the reference field over most of the domain and differ mainly on the steep side of the wake and in the layer immediately adjacent to the sphere.

The angular and centerline profiles also show that the finest Eulerian grid does not always give the smallest near-surface error. For $\Delta=1.0$~mm, the boundary-shell band has a half-width of $\Delta/2=0.5$~mm from the sphere surface. This length scale is comparable to the near-surface resolution of the CFD reference, which resolves the sphere with approximately twelve lattice nodes per diameter. In equivalent physical units, this corresponds to a lattice spacing of roughly $0.5$--$1$~mm. The finest LE-DM shell band is thus close to the resolution limit of the reference field near the sphere. The near-surface comparison at this grid spacing is consequently influenced by the CFD resolution as well as by the reconstruction error. This interpretation is consistent with the wall-region error in Fig.~\ref{fig:cfd_error_sweep}, which saturates rather than continuing to decrease as the grid is refined. It is also consistent with the data-support limitation identified in the analytical sweep of Section~\ref{sec:val_oscsphere}. Away from this thin near-surface band, the bulk wake is recovered faithfully even at the lowest seeding density. The remaining error is concentrated mainly in the steep-gradient region next to the moving body.

The largest errors occur next to the sphere, where the velocity gradients are strongest and the kinematic boundary condition is imposed. This region is also the most sensitive to the finite resolution of the CFD reference and to interpolation error during particle advection. It provides the most stringent part of the test because the synthetic tracks must remain outside the moving body while still sampling the rapid spatial variation of the near-body flow. The particle filtering and corrective near-surface displacement enforce this consistency and preserve the wake and near-sphere structure used for the comparison. The remaining near-body error can then be attributed to the reconstruction rather than to non-physical tracer placement.

The CFD case extends the assessment from the idealized analytical benchmark to a fully three-dimensional rising-sphere flow with a known numerical reference field. The reconstruction error decreases as the grid is refined toward the available tracer spacing and generally improves with increasing tracer density, consistent with the particle-support limitation identified in Section~\ref{sec:val_oscsphere}. The largest errors occur in the wake, where the velocity gradients are steep and the local particle support is less uniform, while the wall-region directional deviation remains small across all configurations. The CFD results therefore provide an intermediate validation step between the noise-free analytical case and the fully experimental RIM-PTV case considered in Section~\ref{sec:val_piv}.

% ---------------------------------------------------------------------
\subsection{Tomographic PTV experiment: freely rising sphere in a refractive-index-matched solution}
\label{sec:val_piv}
% ---------------------------------------------------------------------

The final case applies the framework to real volumetric experimental data, for which the relevant sources of measurement uncertainty are present, including optical distortion, sphere-detection error, occluded tracers, position noise, and a turbulent, three-dimensional wake. Because no independent reference field is available, this case is a demonstration and consistency assessment rather than a quantitative validation. The experimental dataset was acquired in our previous study on refractive-index-matched (RIM) tomographic particle tracking velocimetry of buoyancy-driven spheres~\citep{jose2026application}. We summarize only the elements needed to interpret the present reconstruction. The full experimental protocol and a detailed description of the sphere-detection algorithm are provided in that work.

\begin{figure}
    \centering
    \includegraphics[width=0.95\linewidth]{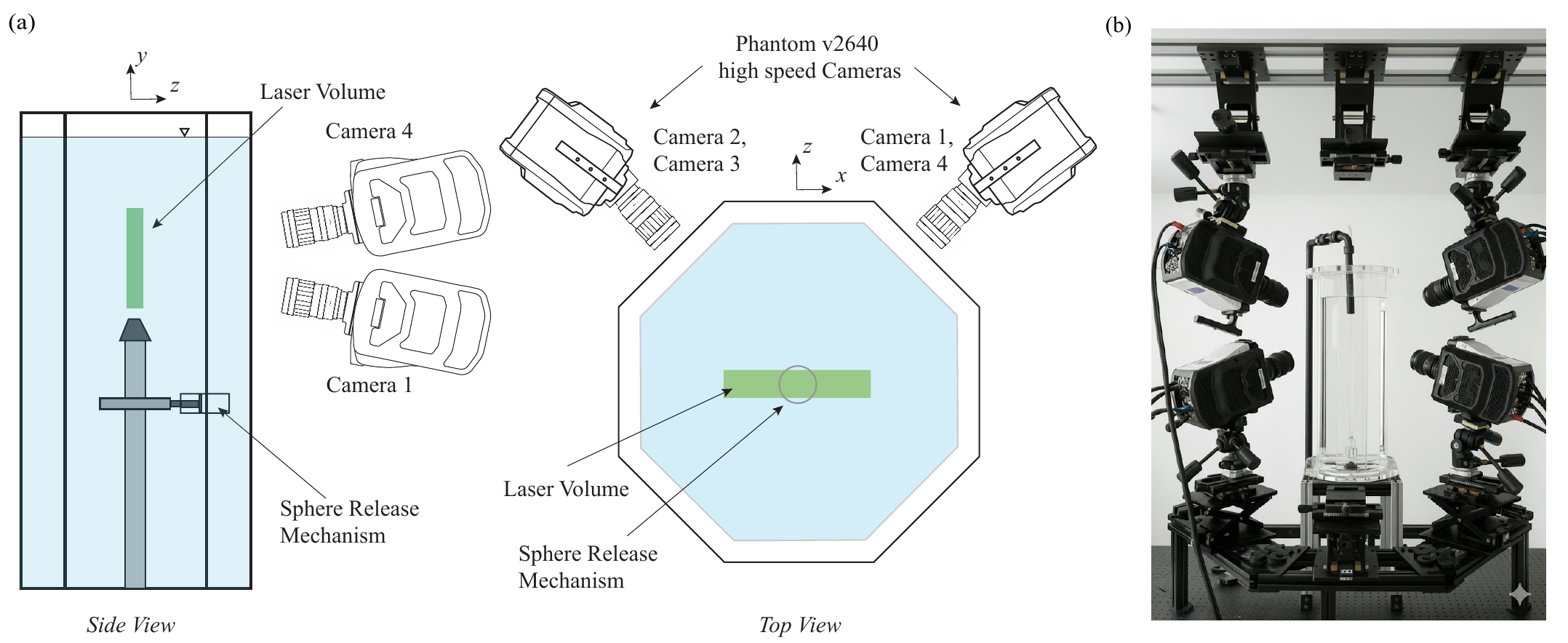}
    \caption{Experimental facility. (a) Side view and top view of the octagonal acrylic tank (110~mm side length, 700~mm height) used for the buoyancy-driven sphere experiment, showing the laser illumination volume, sphere release mechanism, and four Phantom v2640 high-speed cameras. (b) Photograph of the assembled tomographic PTV setup. Adapted from \citet{jose2026application}}
    \label{fig:exptsetup}
\end{figure}

\subsubsection{Experimental configuration.}
A clear acrylic sphere of diameter $d = 11.11$~mm and density ratio $m^* = \rho_s/\rho_f = 0.65$ rises freely through a quiescent aqueous sodium iodide (NaI) solution contained in an octagonal acrylic tank (side length $110$~mm, height $700$~mm) as shown in Fig.~\ref{fig:exptsetup}. The NaI concentration is tuned to $\sim62\%$ by weight to match the refractive index of acrylic, $n = 1.489$, following~\citet{bai2014refractive}. This suppresses optical distortions at the curved fluid-sphere interface and allows tracers in the immediate vicinity of the body to be imaged without aberration. Rhodamine-B-coated PMMA tracers are used together with $550$~nm long-pass filters on each camera, exploiting fluorescent emission to eliminate the residual reflections that would otherwise corrupt near-body particle reconstruction. The consequence of this optical matching is that the sphere itself becomes effectively invisible, and its instantaneous position must be inferred from the surrounding flow signature rather than from direct imaging. Time-resolved tomographic PTV is performed at $4800$~Hz with four Phantom v2640 cameras over a measurement volume of $110 \times 75 \times 25$~mm. At the chosen seeding density of $\sim 0.4$ particles/mm$^3$, approximately $80{,}000$--$100{,}000$ tracers are imaged in the volume. Particle trajectories are obtained from the Shake-the-Box algorithm~\citep{schanz2016shake}, retaining tracks of at least four consecutive frames. The reconstructed Reynolds number based on terminal velocity and diameter is $\mathrm{Re}_T \approx 2880$, and the Galileo number is $\mathrm{Ga} \approx 2397$, placing the sphere near the boundary between the symmetric two-ring (2R) and the asymmetric four-ring (4R) shedding regimes~\citep{horowitz2010effect, Auguste2018}.

\subsubsection{Sphere detection and uncertainty.}
The sphere is localized at every snapshot using the physics-informed detection algorithm of~\citet{jose2026application}, which minimizes a cost function combining void density in the tracer field, vertical-velocity structure, and $Q$-criterion intensity over a candidate region. The detection uncertainty is quantified per snapshot from the local curvature of the cost function and supplies the body-detection uncertainty $\sigma_\Gamma$ that enters the proximity-reweighting term of the LE-DM formulation (Eq.~\ref{eq:reweighting}). For the $d = 11.11$~mm sphere the reported mean positional uncertainty is $0.7\%\,d = 0.078$~mm, with worst-case excursions of $4.0\%\,d = 0.44$~mm. The supplied $\sigma_\Gamma$ is a per-snapshot value in this range, so the proximity reweighting damps near-body tracer contributions on a length scale matched to the actual detection precision rather than to an externally fixed parameter. This is precisely the regime in which the noise-free analytical sweep of Section~\ref{sec:val_oscsphere} indicates that $\sigma_\Gamma$ should become a meaningful parameter.

\subsubsection{Reconstruction setup.}
The LE-DM reconstruction is performed on a uniform Cartesian grid with spacing $\Delta=1$~mm. The grid covers a sphere-centered subvolume that is extended along the rise direction to include the near wake. The rise axis is taken as $+y$ in the laboratory frame. The translational sphere velocity $\mathbf{U}_s(t_k)$ used to construct $\mathbf{u}_\Gamma$ in Eq.~\ref{eq:u_gamma} is obtained by differentiating a low-order polynomial fit to the detected sphere-center positions over the full snapshot sequence. This avoids the velocity noise that would result from frame-to-frame differencing of positions with finite detection uncertainty. The angular velocity $\boldsymbol{\omega}_s(t_k)$ is set to zero because it is not resolved by the present sphere-detection procedure. For the present case, this approximation is expected to have limited influence on the reconstructed near-body field because the imposed boundary motion is dominated by translation. Any rotational contribution would enter through the surface velocity $\boldsymbol{\omega}_s\times a$, which is small compared with the translational component for the observed path-instability regime.

The finite-support interpolation kernel in Eq.~\ref{eq:kernel} and the coverage-adaptive smoothness term in Eq.~\ref{eq:smooth} are both active. The smoothness weight $\lambda_c$ is fixed at the value selected from the synthetic sparse-data benchmark and is kept identical for all experimental reconstructions. The Tikhonov regularization parameter is set to $\kappa=100$. The data-fidelity weights use the per-particle velocity uncertainty, with the characteristic track-velocity standard deviation set to the measured scatter of the PTV velocities, $\sigma_{u,i}=0.03$~m/s. Using the measured scatter prevents the solver from fitting track-level noise, which would otherwise be spread into the quiescent region through the divergence-free constraint. The data weighting is supplemented by a local outlier weight based on the velocity magnitude within a $5$~mm neighborhood. Tracks whose velocity magnitude exceeds the local median by more than five median-absolute-deviation units are down-weighted. At least ten neighbors are required to apply this local criterion. Tracks with fewer neighbors retain full weight. A local criterion is used instead of a global threshold because the wake contains physically elevated velocities that would otherwise be misidentified as outliers when compared with the quiescent-region distribution. This procedure retains genuine wake signal while reducing the influence of locally anomalous tracks, typically reflection ghosts and false STB matches.

The saddle-point system in Eq.~\ref{eq:saddle_point} is solved with MINRES to a relative tolerance of $10^{-6}$. The same experimental dataset was previously reconstructed using VIC\#~\citep{jeon2022fine}, an evolution of the VIC$^{+}$ technique~\citep{schneiders2016dense}, as part of the RIM-PTV study of~\citet{jose2026application}. That reconstruction assimilated particle velocities and material derivatives over a finite time segment. In the absence of an experimental ground truth field, it provides a useful reference for comparing the LE-DM reconstruction with an established particle-assimilation approach.

\begin{figure}
  \centering
  \includegraphics[width=\textwidth]{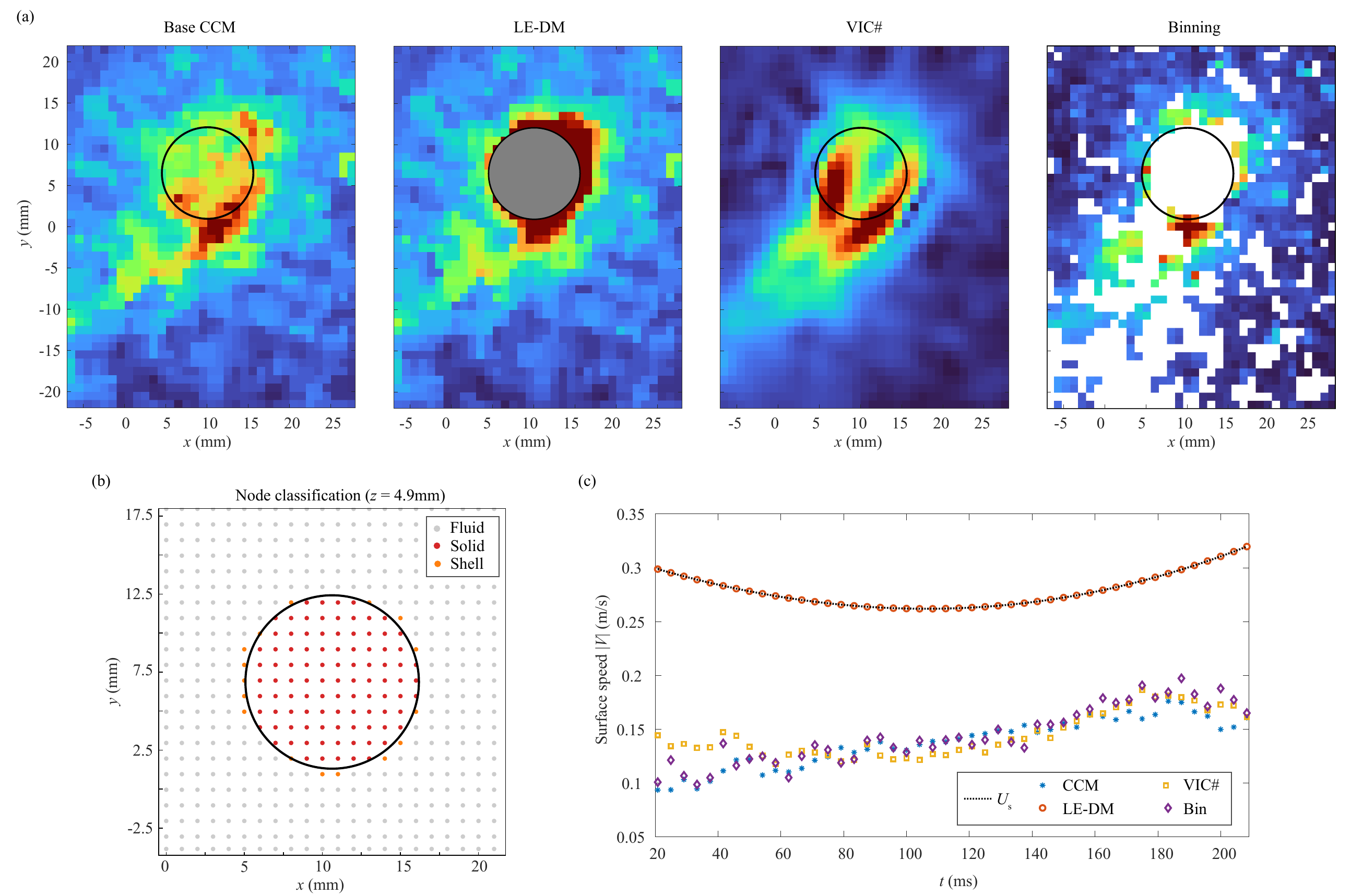}
  \caption{Experimental demonstration of LE-DM on a freely rising rigid sphere ($d = 11.11$~mm, refractive-index-matched solution), at the snapshot $y_s = 6.9$~mm, $|U_s| = 0.264$~m/s. \textbf{(a)} Velocity magnitude on a meridional slice through the sphere center, for the base CCM reconstruction, LE-DM, VIC\#~\citep{jeon2022fine}, and a direct binning of the tracks, all on a common grid and shared color scale, shown without spatial smoothing. Only LE-DM masks the sphere interior (gray disc). The base CCM reconstruction and VIC\# place reconstructed velocity inside the body region, as neither carries a body model, and binning leaves roughly $58\%$ of fluid cells empty (light gray) because naive interpolation cannot close the field where tracks are absent. The white circle marks the detected sphere surface. \textbf{(b)} Node classification produced by the signed-distance function on the same slice, showing solid-interior nodes (set to the body velocity), boundary-shell nodes (where the no-slip condition is enforced), and open-fluid nodes (reconstructed). \textbf{(c)} Mean reconstructed speed over the shell band versus time for all four methods, with the independently measured sphere speed $|U_s(t)|$ as a dashed reference. LE-DM follows $|U_s(t)|$ by construction, since its shell nodes are constrained to the body velocity, while the base CCM reconstruction, VIC\#, and binning lie below it at every snapshot, quantifying how far an unconstrained field departs from the independently measured surface speed.}
  \label{fig:expt_slip_and_wake}
\end{figure}

\subsubsection{Diagnostic comparison with the boundary mask.}

Three diagnostics are used because no independent experimental ground truth is available. The first is the mean boundary-shell mismatch, $\left\langle \left|\mathbf{u}^{\ledm}-\mathbf{u}_\Gamma\right| \right\rangle_{\mathrm{shell}}$, which checks whether the reconstructed velocity on the detected sphere surface matches the prescribed body velocity. Since this condition is imposed as a hard constraint in LE-DM, the quantity should be limited by the solver tolerance. It is used as an implementation check for the boundary block of $\mathbf{B}(t_k)$, rather than as an independent accuracy metric. The second diagnostic is the distance-binned divergence, $\left\langle |\nabla\cdot\mathbf{u}^{\ledm}| \right\rangle_{\phi}$, evaluated over the open-fluid nodes. This verifies that incompressibility is enforced consistently in the bulk and in the near-body region where one-sided stencils are active. A residual that remains at the solver-tolerance level across the distance bins indicates that the stencil assembly and constrained solve are consistent. A localized increase near the interface would instead indicate a stencil or convergence issue. The third diagnostic is a pointwise comparison with the VIC\# reconstruction reported in~\citet{jose2026application}, evaluated on a common grid. This is treated as a cross-method consistency check because both reconstructions start from the same PTV data but use different assimilation principles.

The contribution of the moving-boundary treatment is isolated by repeating the reconstruction with the boundary-mask constraint disabled and enabled, while keeping the particle data and all numerical settings unchanged. The shared settings include the grid, interpolation kernel, $\kappa$, $\lambda_c$, $\sigma_\Gamma$, local-MAD weighting, and MINRES tolerance. The mask-disabled run corresponds to the base CCM reconstruction, while the mask-enabled run corresponds to LE-DM. This comparison is the experimental counterpart of the analytical comparison in Table~\ref{tab:ccm_vs_ledm} of Section~\ref{sec:val_oscsphere}. It isolates the effect of supplying the solid geometry and boundary kinematics to the reconstruction, rather than comparing two independently tuned methods.

Figure~\ref{fig:expt_slip_and_wake} presents the comparison in three parts, with the reconstructed velocity fields in Figure~\ref{fig:expt_slip_and_wake}a, the node classification in Figure~\ref{fig:expt_slip_and_wake}b, and the shell-band speed over the rise in Figure~\ref{fig:expt_slip_and_wake}c. Figure~\ref{fig:expt_slip_and_wake}b shows the node classification produced by the signed-distance function on a meridional slice at one snapshot. The solid-interior nodes are constrained to the body velocity, the boundary-shell nodes enforce the no-slip condition, and the open-fluid nodes are reconstructed from the particle data. This identifies the region where the dynamic mask imposes the kinematic condition in the experimental dataset. At the representative snapshot, the sphere-centered subvolume contains approximately $57{,}000$ grid nodes. Of these nodes, $98.4\%$ are open fluid, while the boundary shell and solid interior account for $0.35\%$ and $1.3\%$, respectively. The constrained nodes are thus confined to the immediate neighborhood of the body. 

The shell-band agreement is primarily an implementation check. For LE-DM, agreement between the reconstructed shell velocity and the body velocity confirms that the detected sphere kinematics have been transferred correctly into the boundary block of the saddle-point system and that the constrained solve has converged. At the representative snapshot used for the field comparison, where the sphere center is at $y_s=6.9$~mm and $|U_s|=0.264$~m/s, the mean shell-band mismatch for LE-DM is $1.8\times10^{-6}$~m/s, which is at the solver tolerance (Table~\ref{tab:expt_crossmethod}). The base CCM reconstruction, which has no solid mask, gives a mismatch of $0.205$~m/s on the same band. VIC\#, which does not impose a no-slip constraint~\citep{jeon2022fine}, gives $0.151$~m/s. These values quantify the departure of unconstrained reconstructions from the known surface velocity on the detected shell.

The first open-fluid cell, one grid spacing from the surface, provides a more direct test of the reconstructed fluid field because its velocity is determined by the particle data and the reconstruction constraints, not prescribed by the boundary condition. All three reconstructions give a data-resolved velocity of order $0.2$~m/s in this region. This is expected because the true near-wall adjustment occurs over a distance smaller than the $1$~mm grid spacing, so the velocity recovers toward the measured near-wall flow within one to two cells. The near-wall fluid slip normalized by $|U_s|$ is comparable across methods, as shown in Table~\ref{tab:expt_crossmethod}. In regions where the sphere is well surrounded by particles, the data already constrain the first-cell velocity and the boundary mask changes it only modestly. The main effect of the mask is to impose the correct surface value and to provide a body-consistent field in the sub-grid region between the detected surface and the first open-fluid cell. This behavior is consistent with a sharp-interface treatment on a fixed Cartesian grid.

Figure~\ref{fig:expt_slip_and_wake}c shows that the shell enforcement is maintained throughout the rise, rather than only at the representative snapshot. The mean reconstructed speed over the shell band is plotted against time for the methods included in the comparison, together with the independently measured sphere speed $|U_s(t)|$ from the detected trajectory. LE-DM follows $|U_s(t)|$ because the shell nodes are constrained to the body velocity. The base CCM reconstruction, VIC\#, and binning remain below the measured sphere speed at every snapshot. Since $|U_s(t)|$ is obtained independently of the velocity reconstruction, this diagnostic shows that methods without a body model do not recover the known surface kinematics. The comparison is limited to the surface-kinematic constraint and does not imply a ranking of open-fluid accuracy.

\begin{table}[ht]
  \centering
  \caption{Cross-method comparison on the experimental rising-sphere
           snapshot ($y_s = 6.9$~mm, $|U_s| = 0.264$~m/s). Shell slip is
           the mean $\|\mathbf{u}-\mathbf{U}_s\|$ over the enforced
           Dirichlet band ($0 < \phi < 0.5$~mm). Near-wall fluid slip is
           the mean over the open-fluid band, normalized by $|U_s|$.
           Quiescent RMS is the root-mean-square speed in the far-field
           region. Wake mean and wake 95th percentile are taken over the
           wake region. No-track fraction is the fraction of wake cells
           without an instantaneous track within one grid spacing, and the
           empty-node fraction is the fraction of fluid cells left
           unfilled by binning after window growth. The raw PTV column
           reports the corresponding statistic of the scattered track
           velocities for reference, and binning is a single-snapshot
           inverse-distance-weighted ensemble average on the same grid.
           All fields are sampled on a common grid.}
  \label{tab:expt_crossmethod}
  \begin{tabular}{lccccc}
    \hline
    Quantity & Raw PTV & Binning & base CCM & LE-DM & VIC\# \\
    \hline
    Shell slip [m/s]               & n/a    & n/a     & $0.205$ & $1.8\times10^{-6}$ & $0.151$ \\
    Near-wall fluid slip / $|U_s|$ & $0.95$ & $0.93$  & $0.97$  & $0.91$  & $0.90$ \\
    Quiescent RMS [m/s]            & $0.039$ & $0.023$ & $0.057$ & $0.057$ & $0.023$ \\
    Wake mean $|V|$ [m/s]          & $0.043$ & $0.033$ & $0.052$ & $0.055$ & $0.034$ \\
    Wake 95th pct $|V|$ [m/s]      & $0.159$ & $0.091$ & n/a     & $0.174$ & $0.132$ \\
    Wake no-track fraction         & $0.34$  & $0.34$  & $0.34$  & $0.34$  & $0.34$ \\
    Empty-node fraction            & n/a     & $0.58$  & n/a     & n/a     & n/a \\
    \hline
  \end{tabular}
\end{table}

\subsubsection{Cross-method consistency in the wake.}

Figure~\ref{fig:expt_slip_and_wake}(a) compares the velocity magnitude on a meridional slice through the sphere center for the base CCM reconstruction, LE-DM, VIC\#, and direct binning of the particle tracks. All fields are resampled onto a common grid and displayed with the same color scale, without additional spatial smoothing. The binning field is an inverse-distance-weighted ensemble average of the tracks within each cell, using the standard DaVis procedure with adaptive window growth in sparsely populated cells. It represents the result obtained from direct track interpolation without imposing physical constraints. The comparison clarifies the role of the reconstruction step. The base CCM reconstruction and LE-DM agree away from the body. VIC\# is smoother because it assimilates information over a finite time window. Direct binning leaves approximately $58\%$ of the wake cells empty even after adaptive window growth, because it cannot close the field in regions with no local tracks. Among the four fields, only LE-DM masks the sphere interior. The base CCM reconstruction and VIC\# assign velocities inside the body because neither method is supplied with the solid geometry.

The quiescent region provides a useful reference for interpreting the differences between methods. In this region, the raw PTV tracks have a velocity scatter of $0.039$~m/s, set by the single-snapshot measurement noise. LE-DM gives a quiescent-region level of $0.057$~m/s, which is consistent with a snapshot reconstruction built directly from noisy tracks. VIC\# gives $0.023$~m/s, below the raw-track scatter, because the finite assimilation window averages over multiple snapshots. The difference between LE-DM and VIC\# in the quiescent region is thus a consequence of snapshot reconstruction versus time-window assimilation, rather than a systematic bias in either method.

All three physics-based reconstructions recover the same wake direction and large-scale structure. With the coverage-adaptive smoothness term active, LE-DM produces a continuous wake field rather than the fragmented field obtained when empty cells are left to the temporal prior alone. At the representative snapshot, the wake has a no-track cell fraction of $0.34$. Over the full $56$-snapshot sequence, this fraction ranges from nearly unity early in the rise, when the sphere has just swept tracers out of its path, to approximately $0.2$ after the wake repopulates. The smoothness term provides the local neighbor-consistency prior needed to close the reconstruction in these temporarily empty cells.

The wake reconstruction is governed primarily by track coverage rather than by the moving-boundary constraint. The base CCM reconstruction and LE-DM agree to within a few percent in median wake amplitude and have identical no-track cell fractions at every snapshot. This shows that the near-body constraints do not materially alter the wake field. The remaining difference relative to VIC\# reflects the reconstruction principle. VIC\# is a time-segment method that fits particle velocity and material acceleration within a vorticity-transport framework over a finite window~\citep{schneiders2016dense,jeon2022fine}. It can use inter-snapshot motion to inform cells that are not occupied by tracks at the current time. LE-DM instead reconstructs each snapshot independently and closes empty cells through the coverage-adaptive smoothness prior. The methods agree where the tracks are dense and differ most in the sparsest part of the wake. This behavior is consistent with the comparison of CCM and VIC\# reported by~\citet{agarwal2021reconstructing}, where the methods give similar results in well-sampled regions and differ near walls and at small scales.

The dependence of the cross-method agreement on track coverage was tested directly with a multiple linear regression. Across the snapshots in which the wake is populated, the ratio of the LE-DM to VIC\# wake-region $99$th-percentile velocity correlates strongly with the wake no-track cell fraction ($R^2 = 0.82$) and shows no significant dependence on sphere height once coverage is accounted for. The apparent decrease in agreement with VIC\# at higher sphere positions is therefore explained by deteriorating track coverage as the sphere rises through the imaging volume, rather than by an intrinsic dependence on trajectory height.

At $\Delta=1$~mm, the saddle-point system is ill-conditioned in the upper-left block because the PTV track density is strongly nonuniform. Shell-region cells carry data weights much larger than those of low-coverage wake cells. The coverage-adaptive smoothness term improves the conditioning by adding spatial coupling at the empty cells that are weakly constrained by the data term. With this term active, the snapshot reconstruction is closed at every fluid node without relying on the temporal prior. The wake field is then determined by the instantaneous tracks and the neighbor-consistency prior, rather than by the initial guess. Re-solving representative snapshots from both a zero initial guess and a warm start gives fields that agree to within the solver tolerance throughout the domain, including the empty wake cells. This confirms that the reported wake field is a well-posed snapshot reconstruction rather than an inherited mode. Wake metrics are reported as regional statistics because they are robust to the residual point-to-point scatter expected in the sparsest cells.

The experimental case demonstrates the complete LE-DM workflow on real tomographic-PTV data, where the reconstruction must operate with measurement noise, nonuniform seeding, near-interface particle loss, and no independent reference velocity field. Under these conditions, the dynamic mask transfers the detected sphere kinematics into the reconstruction to solver tolerance and maintains a divergence-free field over the open-fluid domain. The comparison with the mask-disabled base CCM reconstruction shows that the mask acts locally, at the surface and in the immediately adjacent region. Away from the body, the LE-DM and base CCM fields agree closely because the same particle data and regularization determine the solution.

The surface-kinematic diagnostic provides the clearest distinction between the methods. The base CCM reconstruction, VIC\#, and direct binning all depart from the independently measured sphere speed at the detected surface, whereas LE-DM imposes this velocity by construction. One grid cell away from the surface, the masked and base CCM reconstructions are similar for the present sphere case because the surrounding particle support is sufficient to constrain the near-surface fluid velocity. Thus, in this experiment, the main role of the mask is to provide the correct surface value and a body-consistent sub-grid transition between the detected sphere and the first open-fluid cell, while leaving the wake and bulk reconstruction unchanged where no body is present.  The present experiment exercises LE-DM for a freely translating rigid body in a refractive-index-matched volume. Stationary walls and deforming bodies are limiting cases of the same formulation, as discussed in Section~\ref{sec:generalization}, but they are not tested here. Their reconstruction quality will depend on the availability of near-surface particle data and on the accuracy with which the surface kinematics can be prescribed or measured. Those configurations are left for future experimental validation.

\section{Discussion}
\label{sec:discussion}

LE-DM is most useful where an all-fluid reconstruction is least reliable, in the region at and immediately outside the moving solid surface. This region contains the steepest velocity gradients, the no-slip or rigid-body condition prescribes the admissible surface velocity, and pressure integration is especially sensitive to near-wall errors. An all-fluid reconstruction assigns velocity inside the solid and does not satisfy the surface kinematics, so errors introduced at the interface can propagate into pressure and force estimates that depend on near-wall gradients. LE-DM avoids this inconsistency by carrying both the signed-distance mask and the boundary velocity into the constrained solve. The field is made consistent with the known body motion at the interface, while the data-driven reconstruction is left unchanged in fully fluid regions. When the mask is disabled, the formulation reduces to the base CCM reconstruction. The difference between the two reconstructions throughout this paper is consequently the effect of the mask, which remains localized to the near-interface region because the masking terms are inactive wherever the volume is entirely fluid.

A practical distinction between experimental PTV and CFD-based immersed-boundary calculations is that the geometry and its motion are not known exactly. The detected surface position carries uncertainty, and the particle field near the surface is affected by reflections, optical gradients, occlusion, and particle drop-out. LE-DM accounts for this through the body-detection uncertainty $\sigma_{\Gamma}$, which sets the length scale for damping near-interface measurements. The value used here is the scalar, per-snapshot uncertainty reported by the detection algorithm of \citet{jose2026application}. It is treated as isotropic and allowed to vary in time. This weighting reduces the influence of tracks whose position relative to the detected surface is uncertain. In the clean analytical case, it has little effect on the bulk reconstruction, as shown by the sweep in Section~\ref{sec:val_oscsphere}. It becomes important when $\sigma_{\Gamma}$ represents genuine body-detection uncertainty in the experiment.

A related issue is the loss of particle support in the wake. The finite-support kernel and the coverage-adaptive smoothness term address this by spreading information over a local neighborhood and supplying a neighbor-consistency prior where tracks are absent. These terms regularize the reconstruction, but they do not replace missing measurements, so flow structure smaller than the local track spacing cannot be recovered uniquely. Both the smoothness weight $\lambda_c$ and the damping length $\sigma_{\Gamma}$ are fixed by prescribed rules rather than tuned for each case. The value of $\lambda_c$ is selected once from a synthetic benchmark, and $\sigma_{\Gamma}$ is taken directly from the detector. The reconstructed wake amplitude is therefore an output of the method rather than a fitted quantity. Time-segment assimilation methods such as VIC\# draw on information across frames to fill empty regions \citep{jeon2022fine,scarano2022dense}, whereas LE-DM reconstructs each snapshot independently. The boundary-aware domain and kinematic constraints introduced here are defined per snapshot and do not depend on the temporal treatment.

Several limitations bound the present formulation. The boundary is represented on a fixed Cartesian grid, so the imposed boundary condition is located only to within the grid spacing. Flow features thinner than a cell cannot be resolved by the reconstruction alone, as in any sharp-interface immersed-boundary method on a non-body-fitted grid. The temporal-prior correction at newly exposed nodes uses the boundary velocity as a regularization reference. In regions with very weak particle support, this can bias the field toward the body motion over the first few grid layers. The most important limitation is that the shell kinematics enter as a hard constraint. A noisy or biased boundary velocity, whether from a noisy trajectory, unresolved rotation, or an inaccurate deforming-surface estimate, is transferred directly into the near-wall field. The body detection and surface-velocity estimate are therefore part of the reconstruction error budget. This is why the surface velocity in the present experiment is obtained from a smoothed trajectory fit rather than from frame-to-frame differencing.

The LE-DM formulation is independent of geometry or motion. Stationary walls, rotating or multiple bodies, and deforming bodies can be treated by updating the signed-distance field and prescribing the corresponding surface velocity. For a deforming body, the surface velocity would be defined on the moving interface and extended into the boundary shell, for example by constant-normal extension \citep{sethian1999level}. The node classification, discrete stencils, and constrained solve would otherwise remain unchanged. These cases are not demonstrated here because comparable high-fidelity three-dimensional PTV data are not available. Acquiring such data around a freely moving solid is itself a central experimental challenge because the body limits optical access near the surface, its position and kinematics must be recovered from the measurement, and the wake is fully three-dimensional and unsteady. Refractive-index matching helps overcome these limitations by reducing optical distortion at the curved interface and enabling particle imaging close to the body. The freely rising sphere is therefore a stringent demonstration case rather than a simplified one.

\section{Conclusions}
\label{sec:conclusions}

Lagrangian-to-Eulerian reconstruction strongly influences the accuracy of quantities derived from volumetric PTV, including velocity gradients, pressure, and hydrodynamic loading. Existing constrained reconstruction methods can impose no slip on fixed boundaries, but they do not generally address a solid body that moves through a fixed measurement grid and changes the fluid-solid partition from one snapshot to the next. LE-DM addresses this moving-boundary problem by embedding the body geometry and kinematics directly into the Eulerian reconstruction. At each time step, a signed-distance field classifies grid nodes as open fluid, boundary shell, or solid interior, and the data fit, incompressibility constraint, and surface-kinematic constraint are assembled on the resulting masked domain within a single constrained solve. The reconstructed field is divergence-free in the fluid and consistent with the prescribed body motion at the surface, without requiring a body-fitted mesh, a separate projection step, or a post-solve correction. Stationary boundaries are recovered as the limiting case of a time-independent mask. Since LE-DM operates on tracks already produced by STB or a related method, it can also be used downstream of object-aware tracking methods that improve the Lagrangian data at the image-reconstruction stage.

The method was assessed for a spherical moving boundary through a sequence of analytical, numerical, and experimental cases. The analytical oscillating-sphere case isolates the moving-boundary treatment and recovers the imposed surface motion with the correct phase and direction, with errors that scale consistently with the Stokes-layer thickness and grid spacing. The CFD synthetic-track case extends the assessment to a three-dimensional unsteady wake with a known reference field, where the largest errors occur in the steep-gradient near-body and wake regions. The refractive-index-matched experiment demonstrates the complete workflow on real tomographic-PTV data with no independent reference field. In that case, LE-DM transfers the independently measured sphere kinematics into the reconstructed field to solver tolerance, while the base CCM reconstruction, VIC\#, and direct binning do not recover the known surface speed. Away from the body, LE-DM and the base CCM reconstruction agree closely, confirming that the mask acts locally at the interface.

These results establish LE-DM for a freely translating spherical body and provide a foundation for applying the same formulation to more complex geometries and motions. By enforcing the surface kinematics and removing non-physical solid-interior velocities, LE-DM provides a boundary-consistent velocity field for pressure and force estimation. More broadly, the method makes the moving solid part of the reconstruction problem rather than a region to be masked or corrected after the fact. The measured tracks determine the fluid field, while the signed-distance mask and body kinematics define the admissible domain and surface motion, yielding an Eulerian reconstruction that respects both the particle data and the moving solid boundary.

\section*{Declarations}
\subsubsection{Funding}
This work was partially funded by ACS-PRF-ND grant \#65901-ND9.
\subsubsection{Competing interests}
The authors declare no competing interests.
\subsubsection{Data availability}
The experimental dataset analyzed in this study was acquired in a previous work and is described in \citet{jose2026application}. The reconstruction data and processed fields are available from the corresponding author on reasonable request.

% ============================================================
%  Bibliography  (author-year)
%  For numbered references, use:  \bibliographystyle{unsrtnat}
% ============================================================
\bibliographystyle{plainnat}
\bibliography{references}

\end{document}